\documentclass{aa}
\usepackage{graphicx}

\newcommand{\Teff}{\mbox{$T_{\rm eff}$}}
\newcommand{\Lg}{\mbox{$\log g$}}

\newcommand{\Deg}{{^{\rm o}}}
\newcommand{\Sec}{{\rm s}}
\newcommand{\Kel}{{\rm K}}
\newcommand{\p}{{\frac{\delta P}{P}}}
\newcommand{\lnP}{{\tilde{\omega}}}
\newcommand{\nabad}{{\nabla_{\rm ad}}}
\newcommand{\LegP}{{\rm \bf P}}

\newcommand{\flux}{{\cal F}}
\newcommand{\spec}{{\cal Q}}
\newcommand{\rtg}{{\bf \Omega}}
\newcommand{\nvec}{{\bf n}}
\newcommand{\rvec}{{\bf r}}
\newcommand{\dIdF}{{ \left. \frac{\partial I_{\lambda}}{\partial F} \right|_{F_0}}}
\newcommand{\I}{{\cal I}}

\begin{document}

\title{Nonlinear effects in time--resolved spectra of DAVs}
\author{ J. Ising \and D. Koester}

\institute{Institut f\"ur Theoretische Physik und Astrophysik, Universit\"at
Kiel, D-24098 Kiel, Germany (ising,koester@astrophysik.uni-kiel.de)}

\offprints{D. Koester}

\date{\today}


\abstract{
Numerical simulations of light curves of variable DA white dwarfs (ZZ
Ceti stars) predict flux amplitudes with surface distributions different
from the spherical harmonics of the pulsation mode in deeper layers. In
contrast to the results of the perturbation analysis by Gold\-reich and Wu
this is also true for the fundamental period of the flux variation. As a
consequence normalized amplitude spectra depend not only on the mode
number $l$ but also on pulsation amplitude and inclination. Another new
result is that with increasing amplitude of the pressure variation below
the convection zone the flux variation at the surface goes through a
maximum and then decreases again.
}

\maketitle

\section{Introduction}
The ZZ~Ceti stars are the coolest class of pulsating white
dwarfs. They are multi-periodic $g$-mode pulsators with periods in the
range of $100\Sec$ to $1000\Sec$. In contrast to the PG~1159 objects
and the variable DB only a few modes are observed in each individual
ZZ~Ceti star. The number of observed pulsation periods, the amplitudes
and the shape of the light curves of these stars change systematically
within the instability strip (see e.g. Clemens
\cite{Clem93}). At the hot (blue) edge small amplitude
pulsations are observed with sinusoidal light curves. The cooler
ZZ~Ceti stars show more modes with large amplitudes of more than
$10\%$ with overtones and additional frequencies, formed by sums and
differences of the fundamental frequencies of the modes.

The small number of unstable modes forms an obstacle to standard mode
identification techniques (identifying the ``quantum numbers'' $l$ and
$m$ of the spherical harmonics of a mode), which rely mostly on
periods and period spacings of unstable frequencies in the power
spectrum.  In recent years time--resolved spectroscopy of white dwarfs
has become possible and was used to identify the spherical harmonics
of the modes.  This mode identification technique uses the wavelength
dependence of normalized pulsation amplitudes and takes advantage of
the wavelength dependence of the limb darkening and different
cancellation of the flux variation for different spherical harmonics
(Robinson et al. \cite{Robi82}; Brassard et al. \cite{Bras95};
Robinson et al. \cite{Robi95}; Kepler et al. \cite{Kep00}; Clemens and
Kerkwijk \cite{Clem00}). This diagnostic technique depends on the
assumption that the flux amplitude on the visible surface varies with
the spherical harmonic ($l, m$) of the mode.

This assumption --- which we will call ``linear'' -- is subject
to criticism, at least for larger amplitudes.  For large amplitudes
non--sinusoidal light curves are observed, in contrast to the
sinusoidal light curves for small amplitudes. Because the ZZ~Ceti
stars are non-radial pulsators, the amplitude varies over the surface
of the star and one would expect a variation of the shape of the light
curves over the surface for large amplitudes as well. This is in
contradiction to the assumption of a flux variation with spherical
harmonics.  
For a quantitative description of the integrated light curve and
spectra we will need a theoretical understanding of the reaction of
the local surface flux to an imposed pressure variation at deeper
layers. This theory should demonstrate how the surface variation
becomes increasingly non--sinusoidal with increasing amplitude of the
pressure variation.

Such a description was provided by Brickhill (1983), who
proposed, that the thin convection zone below the surface of the
ZZ~Ceti stars has an important effect on the light curves. Brickhill
showed in a series of papers, that numerical simulations of the light
curves can qualitatively reproduce the observed flux variations.  This
result was confirmed by Wu (\cite{Wu98}) and Wu (\cite{Wu00}).

Goldreich \& Wu (\cite{Gold99}, \cite{Gold99a}) and Wu \& Goldreich
(\cite{Wu99}) studied the driving mechanism of the ZZ~Ceti stars in a
linear perturbation analysis. Their work lead to an impressive
progress in the understanding of the driving of the pulsation modes,
but it their linear analysis they consider only the entropy change in
a convection zone of constant depth. Wu (\cite{Wu98},\cite{Wu00})
extended this work to take into account the depth changes, but used
only static models of convection zone.

The numerical integrations we have performed for this paper show that
for larger amplitudes the depth varies significantly during one
cycle. At each time step during the pulsation the structure remains
different from any static structure, leading to modifications of the
nonlinear effects, which appear already in the perturbation analysis
using static structures, as described in this paper.

We extend Brickhill's work and show that the convection
zone strongly influences the spatial distribution of surface
amplitudes, which for larger amplitudes cannot anymore be
described by spherical harmonics.  This cannot be neglected in the
technique of mode identification by time--resolved spectroscopy.
In the next section we describe the model used for our numerical
calculations.  First we will introduce the relatively simple system of
ordinary differential equations to solve the time--dependent structure
of the envelope of ZZ~Ceti stars, obtained by allowing only a
restricted time dependence of the basic equations. To examine the
validity of this restriction for large amplitude pulsation we
introduce the full time--dependent equations in the following
subsection. The boundary conditions have a large influence on the
results of the numerical simulations; the following subsection is
devoted to this topic. The 3. section describes the results of a light
curve simulation for a plane--parallel column with a fixed pulsation
amplitude. In the 4. section we consider that the pulsation amplitude
varies over the surface for non-radial pulsators like the ZZ~Ceti
stars, and describe the surface distribution of the flux for different
maximum amplitudes and spherical harmonics. The technique used to
calculate the total flux together with the expected consequences for
the time--dependent spectroscopy are discussed in the
5. section. Finally we give our main conclusion in section 6.

\section{The numerical model}
Within her analysis Wu (\cite{Wu98},\cite{Wu00} calculates the local
response of the surface flux to a sinusoidal variation of pressure and
flux in a layer at the bottom of the convection zone.  Due to the
change in extent and structure of the convection zone its reaction to
the perturbation of the input flux is nonlinear, leading to the
appearance of higher harmonics in the surface flux (Wu \cite{Wu00})).

The main result is that the amplitude of the fundamental period
depends linearly and that of the first overtone quadratically on the
amplitude below the convection zone. The phase delays between surface
flux and pressure variation are constant, independent of amplitude.
This important result guarantees that the spatial distribution over
the stellar surface for the fundamental period is still given by the
same spherical harmonic, which describes the variation in the deeper
layers. This justifies the usual assumptions made in the analysis of
time--dependent spectroscopy.

The starting point of our own study is the numerical model of
Brickhill.  In this model the convective layer is considered as a
plane--parallel column.  The lower boundary of this column lies below
the surface convection zone, the upper boundary is directly below the
photosphere. The sinusoidal relative pressure variation is assumed to
be constant through the whole column, an assumption justified in the
work by Brickhill and Goldreich and Wu.

In the simulation we calculate the transfer of energy and the
time--dependent temperature structure numerically in a way similar to
Brickhill's work. Some improvements over his work are state-of-the-art
equation of state from Saumon et al.(\cite{Saum95}) and OPAL opacity
data (Iglesias and Rogers, \cite{Igle96}), and our own model
atmosphere grids, which are described in Finley et al. (\cite{Fin97}).

In this paper we want to show the principal effects for the analysis
of time--resolved spectroscopy. Since we are not aiming at a detailed
comparison with observations, it is sufficient to investigate only one
equilibrium model with one effective temperature and one pulsation
period. We take a stellar model with $\Teff=11350\Kel$ and $\log g =
8$ and a pulsation period of $100\Sec$. We use only one parameter for
the mixing length theory (ML2/$\alpha=0.6$); the thermal
relaxation time scale is 4.5~s. With this choice our model is
representative for the conditions near the blue edge of the
instability strip. All our calculations and results of this study are
based on this set of parameters.

\subsection{The basic equations}
The main aim of the model calculation is to determine the temperature
structure as a function of time. The tempe\-rature disturbance $\delta
T$ for a static model can be calculated from
\begin{equation} \label{eqn1}
\delta T = \frac{\delta Q}{C_p \Delta M}+\nabla_{ad} T \p .
\end{equation}
Equation~(\ref{eqn1}) is equivalent to the first law of thermodynamics
(e.g. Cox \cite{Cox80}, p. 37) and can be found in Brickhill
(\cite{Brick83}) as well. The first term on the right describes the
non--adiabatic heating by the additional heat energy $\delta Q$ of a
mass $\Delta M$ with the heat capacity $C_p$. The second term is the
adiabatic contribution to the temperature change due to the relative
pressure variation $\p$ in an environment with an adiabatic
temperature gradient $\nabla_{ad}$.

To calculate the heat energy $\delta Q$ we consider a finite volume
element with the cross section $A$ and a height $\Delta
z$. Introducing a discretization and finite volume elements at this
point, we can avoid the use of partial differential equations. In a
plane--parallel symmetry the flux has only a vertical component. If
$\Delta F$ is the difference between the outgoing and the incoming
flux for a mass element, the heat energy accumulated in a time
interval $\Delta t$ is given by
\begin{equation} \label{eqn2}
\delta Q = \int^{\Delta t}_0 dt \, A \Delta F
\end{equation}
or written in differential form
\begin{equation} \label{eqn3}
\frac{d \delta Q}{dt} = A \Delta F
\end{equation}

To calculate the mass $\Delta M$ of the volume element we follow the
procedure of Brickhill.  We take the pressure of the non--pulsating
star in hydrostatic equilibrium to define the vertical depth
scale. Following Brickhill we introduce $\lnP \equiv \ln P$ as the
independent variable of the problem. The mass element for a fixed
pressure difference $\Delta \lnP$ can be written as
\begin{eqnarray}
\Delta M &=& A \int_{z_1}^{z_2} dz \, \varrho(t) \nonumber \\
& = &A \int_{\lnP_1+\Delta \lnP}^{\lnP_1}
d\lnP \, \varrho(t) \frac{dz}{d\lnP} = A \int_{\lnP_1}^{\lnP_1+\Delta \lnP}
d\lnP \,
\frac{P(t)}{g}. \label{eqn4}
\end{eqnarray}
As Brickhill assumed and was confirmed by Goldreich \& Wu
(\cite{Gold99}) the relative pressure variation can be taken as
constant through the whole column. We further assume a constant
surface gravity $g$ throughout the column. This simplifies the
expression for the mass element:
\begin{eqnarray}
\Delta M &=& A \frac{\exp(\p(t))}{g} \exp(\lnP) \int_0^{\Delta \lnP} d\lnP' \,
\exp(\lnP') \nonumber\\
& = & A \frac{P(t)}{g} \hat{f}(\Delta \lnP) \label{eqn5}
\end{eqnarray}
with
\[ \hat{f}(x) \equiv \exp(x)-1. \]

In a first order approach we can combine equation (\ref{eqn1}),
(\ref{eqn3}) and (\ref{eqn4}), by taking the derivative with respect
to the time of (\ref{eqn1}). In this first order approach we are only
concerned with the time--dependency of $\delta T$, $\delta Q$ and
$\p$. We get
\begin{equation} \label{eqn9}
\frac{dT}{dt} = \frac{\Delta F \, g}{C_p P \hat{f}(\Delta \lnP)} + 
\nabla_{ad} T \frac{d}{dt} \p.
\end{equation}
For all the mass elements of the vertical column, equation
(\ref{eqn9}) defines a system of ordinary differential equations (ODE)
for the temperature structure. The fluxes describe the coupling of the
equations at each grid-point of the column. In our approximation we
calculate the flux locally from the actual temperatures and
temperature gradients. It is given by
\begin{equation} \label{eqn11}
F = F_{\rm r} + F_{\rm c},
\end{equation}
where the radiative flux $F_{\rm r}$ can be calculated with the
diffusion approximation and the convective flux $F_{\rm c}$ by means
of the mixing length theory. We use a time--independent formulation of
the mixing length theory and assume an instantaneous adjustment of the
total flux to the temperature and pressure structure of the column. To
check the validity of this assumption we have also used the
time--dependent Kuhfuss model for convection (Kuhfuss, \cite{Kuhf86};
Wuchterl \& Feuchtinger, \cite{Wucht98}) and found no differences
between the time--dependent and the time--independent model
calculations. This result holds strictly, if the kinetic energy of the
convection does not contribute to the thermal energy, but the
differences are also small, if we take such a coupling into account.

The rapid variation of the flux with the temperature leads to short
relaxation time scales near the upper boundary of the column.  This
leads to a stiff behavior of the ODE. Consequently we use a Gear
solver (Gear, \cite{Gear71}) to integrate equation
(\ref{eqn9}).

\subsection{The full time--dependent equations}
The differential formulation of (\ref{eqn1}) leads in principle to
reliable solutions for large perturbations as well. But to obtain
equation (\ref{eqn9}) we have ignored all time--dependencies except for
$\delta T$, $\delta Q$ and $\p$. For large amplitudes ($\p > 5\%$)
this cannot be justified any longer, since temperature and specific
heat capacity can change by large amounts. In a first step to correct
for this we can replace all quantities in equation (\ref{eqn9}) by
their current values instead of those of the equilibrium model. Since
this may not be sufficient for short pulsation periods ($\sim
100\Sec$), we have modified equation (\ref{eqn9}) and taken into
account all terms produced by the derivative of (\ref{eqn1}) with
respect to time
\begin{equation} \label{eqn9a}
\left[1+x_1-x_2\right] \frac{dT}{dt} = \frac{\Delta F \, g}{C_p P 
\hat{f}(\Delta \lnP)} + \nabla_{ad} T \frac{d}{dt} \p - x_3 + x_4
\end{equation}

In the following all thermodynamic quantities are functions of $T$ and
$P$, and a partial derivative with respect to $T$ means that $P$ is to
be kept constant and vice versa.
The correction terms $x_1$ and $x_3$ due to the non--adiabatic term of
equation (\ref{eqn1}) are
\begin{eqnarray*}
x_1 & = & \frac{\delta Q}{C_p^2 \Delta M} \frac{\partial C_p}{\partial T} \\
x_3 & = & \frac{\delta Q}{C_p^2 \Delta M} \frac{\partial C_p}{\partial P}
\frac{\partial P}{\partial t}
\end{eqnarray*}
The correction terms $x_2$ and $x_4$ are produced by the derivative of the
adiabatic term of (\ref{eqn1})
\begin{eqnarray*}
x_2 & = & \left(\frac{\partial \nabad}{\partial T} T + \nabad \right) \p \\
x_4 & = & \frac{\partial \nabad}{\partial P} T \p 
\frac{\partial P}{\partial t}
\end{eqnarray*}

The derivative of the mass element with respect to the time is
calculated again under the assumption, that the relative pressure
perturbation is constant through the whole column. Then this variation
is inversely related to the change of the horizontal area
\[ \frac{\delta A}{A} = - \p. \]
This leads to
\begin{equation} \label{eqn8}
\frac{d\Delta M}{dt} = \left(\frac{dA}{dt} P(t) + A(t) \frac{dP}{dt} \right)
\frac{\hat{f}(\Delta \lnP)}{g} = 0.
\end{equation}

In equation (\ref{eqn9a}) appear both the time--derivative of $\delta
Q$ and $\delta Q$ itself. This changes the structure of the equations
from ordinary differential equations to a set of integro--differential
equations.  We solve this system by a splitting procedure. In the
inner step we solve the differential equations for a time step short
compared to the pulsation period with constant $\delta Q$. After this
time step $\delta Q$ is updated.  For time steps of about $0.25\%$ of
the period the achieved accuracy is high enough, since the integral
only appears in higher oder terms.

The contributions of $x_1$ and $x_2$ are important for large pulsation
amplitudes, caused by the large and rapid changing of the temperature,
whereas all terms including a derivative of $\nabad$ are unimportant.
Table (\ref{tab1}) compares the amplitudes of the fundamental and the
first overtone of the photospheric flux variation as a result of
(\ref{eqn9}) with (\ref{eqn9a}) for our model with a period of $100
\Sec$. For (\ref{eqn9}) we take the current values of all quantities
instead of the values of the equilibrium model.
\begin{table}[htb]
{\begin{center}
\begin{tabular}{|c|c|c|c|c|}
\hline
& \multicolumn{2}{|c|}{ODE} & \multicolumn{2}{|c|}{IDE} \\ \hline
$\p$ & $A_1$[\%] & $A_2$[\%] & $A_1$[\%] & $A_2$[\%] \\ \hline
$0.02$ & $1.71$ & $0.24$ & $1.71$ & $0.25$ \\
$0.04$ & $3.64$ & $1.17$ & $3.65$ & $1.23$ \\
$0.06$ & $5.14$ & $2.01$ & $5.18$ & $2.13$ \\
$0.08$ & $6.17$ & $2.61$ & $6.29$ & $2.83$ \\
$0.10$ & $6.78$ & $2.95$ & $7.04$ & $3.32$ \\
$0.12$ & $6.97$ & $2.97$ & $7.45$ & $3.56$ \\
$0.14$ & $6.74$ & $2.67$ & $7.54$ & $3.54$ \\
$0.16$ & $6.09$ & $2.01$ & $7.31$ & $3.25$ \\
$0.18$ & $5.17$ & $1.17$ & $6.78$ & $2.68$ \\ \hline
\end{tabular}
\end{center}}
\caption{Fourier amplitudes for the photospheric flux
variation. Column 2 and 4 display the amplitude of the fundamental for
equation (\ref{eqn9}) (ODE) and equation (\ref{eqn9a}) (IDE),
respectively. Column 3 and 5 show the amplitudes of the first
overtone. All quantities are functions of the pressure amplitude
$\p$.}
\label{tab1}
\end{table}

The differences for large pressure amplitudes are significant in the
solutions of (\ref{eqn9}) and (\ref{eqn9a}), but the qualitative
behavior is the same in both cases. Since the numerical effort is not
very much larger for solving the IDE this method will be implemented
in future versions of our code. The ODE solution (\ref{eqn9})
is adequate for a qualitative discussion and all results given in
this paper are based on (\ref{eqn9}) with current values for all
quantities. One should remember, however, that the results for large
amplitudes --- though qualitatively correct --- may be affected by
this approximation.

\subsection{Boundary conditions}
The basic system of ODE (\ref{eqn9}) looks like an initial value
problem, but this is only partially correct. To obtain equation
(\ref{eqn2}) we have introduced a discretization of the column, to
replace the spatial derivatives of the flux with finite differences.
The mathematical structure of a partial differential equation survives
in the coupling of the ODEs and the necessity to define boundary
conditions for the flux.

For the lower boundary we assume the adiabatic approximation. The
matter in the region below the convection zone is well approximated as
a completely ionized hydrogen plasma. Under this assumption the flux
perturbation is given by (see Goldreich \& Wu, \cite{Gold99})
\begin{equation} \label{eqn10}
\frac{\delta F}{F} = \frac{3-3\kappa_\varrho-2\kappa_T}{5} \p,
\end{equation}
where $\kappa_\varrho$ and $\kappa_T$ are the logarithmic derivatives
of the opacity with respect to the density and temperature.  This
adiabatic assumption is the weakest point of the approximations in the
numerical model. In order to test this assumption and also to find the
optimum location of the lower boundary we have made a numerical
experiment.

\begin{figure}[htb]
\includegraphics*[width=8.8cm]{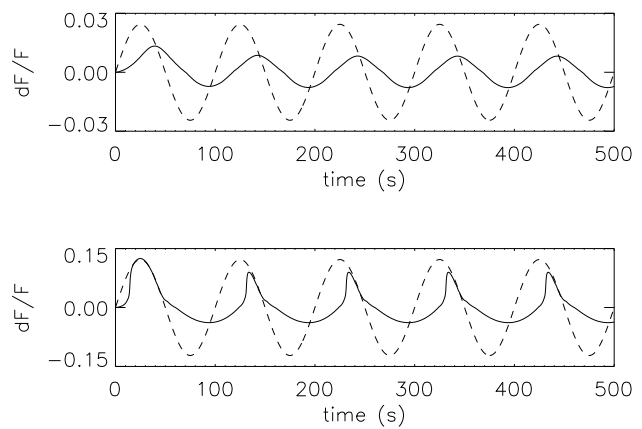}
\caption{
The variation of the photospheric flux (solid line) and the flux variation
at the lower boundary (dashed line) are plotted for a $1\%$ (upper panel)
and a $5\%$ pressure amplitude (lower panel).}
\label{Fig1} 
\end{figure}

We assume a column with a lower boundary much deeper than the bottom
of the convection zone with a constant pressure amplitude in the whole
column. The depth of the lower boundary is limited by the growing
computing costs, because a pulsation simulation has to be done for at
least one thermal time scale of the whole column to reach a stationary
situation. For our model we choose a thermal time scale of $1500\Sec$
and a value of $\lnP = 22$ to fix the lower boundary for this
experiment. At this lower boundary we apply a sinusoidal flux
perturbation and study the resulting flux at a point clearly below the
maximum extent of the convection zone, but far away from the lower
boundary. At this point the flux is no longer sinusoidal but shows a
fundamental with amplitude $A_1$ and a first overtone with amplitude
$A_2$. The phases are shifted relative to the bottom flux by $\phi_1$
and $\phi_2$ respectively, where a positive phase means that the
variation is lagging behind the perturbation. One should note that
these effects would be absent in a completely radiative star; they are
caused by the delayed reaction of the convection zone, which
determines the outer boundary for the envelope structure.

Table~\ref{tab2} shows the results at the point with $\lnP =
19.45$. The first overtone at this point remains at least one order of
magnitude smaller than the fundamental amplitude and the phase shift
between pressure and flux variation remains small. We follow Brickhill
in our conclusion that the adiabatic flux variation at the lower
boundary is a good approximation if the propagation zone of the
$g$-modes is far away from the convective layer. This is true for
short period pulsations as considered in our model, which represents
the ``blue edge'' of the instability strip. For our pulsation
simulations we then fixed our lower boundary at $\lnP=20$.
\begin{table}[htb]
\begin{center}
\begin{tabular}{|c|c|c|c|c|c|}
\hline
$\p$ & $\left(\frac{\delta F}{F}\right)_{\rm lb}$[\%] & $A_1$[\%] & 
$\phi_1[\Deg]$ & $A_2$[\%] & $\phi_2[\Deg]$ \\
\hline
$0.02$ & $4.40$  & $4.72$  & $2.67$ & $0.05$ & $-98.43$ \\ 
$0.04$ & $8.80$  & $9.43$  & $2.67$ & $0.19$ & $-97.35$ \\
$0.06$ & $13.20$ & $14.09$ & $2.84$ & $0.40$ & $-97.80$ \\
$0.08$ & $17.60$ & $18.62$ & $3.12$ & $0.69$ & $-97.80$ \\
$0.10$ & $22.00$ & $23.05$ & $3.43$ & $1.08$ & $-97.52$ \\
$0.12$ & $26.40$ & $27.39$ & $3.75$ & $1.56$ & $-97.35$ \\
$0.14$ & $30.80$ & $31.59$ & $4.09$ & $2.10$ & $-97.92$ \\
$0.16$ & $35.20$ & $35.61$ & $4.47$ & $2.72$ & $-99.01$ \\
$0.18$ & $39.60$ & $39.45$ & $4.88$ & $3.44$ & $-100.32$ \\
\hline
\end{tabular}
\end{center}
\caption{Influence of non--adiabatic terms at the lower boundary:
Results of the experiment described in the text. The $2^{\rm nd}$
column displays the amplitude of the sinusoidal flux variation at the
lower boundary. The $3^{\rm rd}$ and $4^{\rm th}$ column is the
amplitude and phase of the fundamental at the point $\lnP = 19.45$,
which is always below the convection zone. The first overtones are
shown in column $5$ and $6$.}
\label{tab2}
\end{table}

To find an upper boundary condition we make the assumption, that the
photosphere reaches the static structure for each time step
instantaneously.  As discussed e.g. by Brickhill (\cite{Brick83}) this
approximation is valid for small thermal time scales compared with the
pulsation period. We choose the upper boundary at the pressure point
with the optical depth $\approx 10$, where this condition is clearly
fulfilled. The advantage to take the optical depth clearly greater
than $1$ at the upper boundary of the column is, that the
time--dependent calculations can be done with the diffusion
approximation for the radiative flux. The determination of the
photospheric structure requires a detailed radiative transfer
calculation, but without consideration of terms with
time--derivatives. To connect these two different calculations we
assume, that the combination of the actual flux, pressure and
temperature is the same at the upper boundary of the column as in one
specific static photospheric model. To find the flux we use a grid of
static models and find the known combination of pressure and
temperature in this grid for each time step separately and interpolate
the flux from these models. This procedure avoids a further
linearization at the upper boundary, but is in principle equivalent to
Brickhill's method.

\begin{figure}[htb]
\includegraphics*[width=8.8cm]{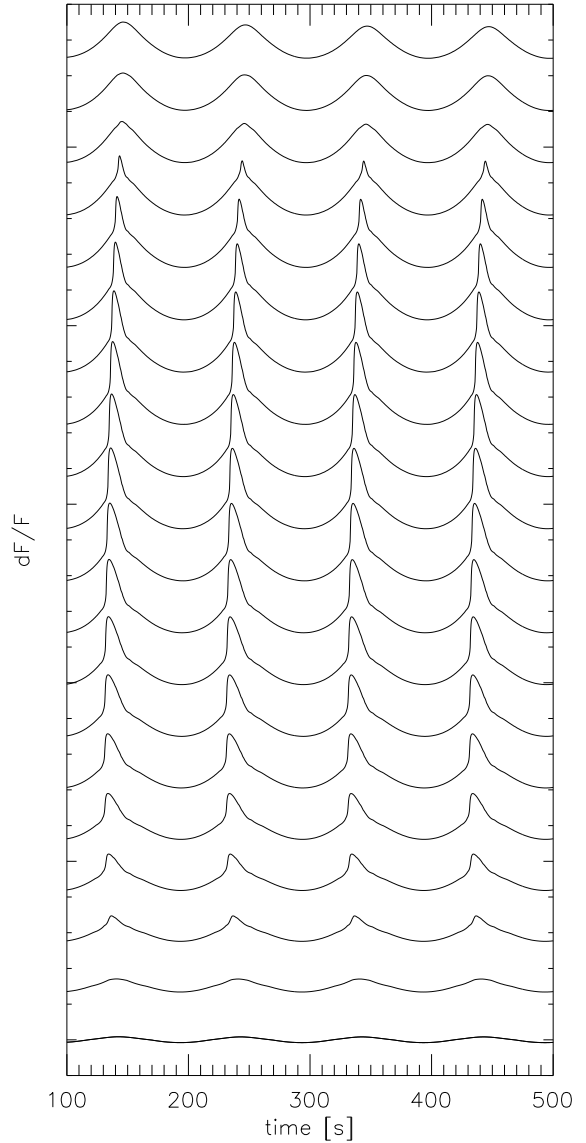}
\caption{ Shape of the flux variation for different pressure
amplitudes.  The bottom line is the light curve for $1\%$ amplitude,
the increment between curves is $1\%$. The light curve at the top of
the panel is for $20\%$ pressure amplitude.}
\label{Fig2}
\end{figure}

\begin{figure}[htb]
\includegraphics*[width=8.8cm]{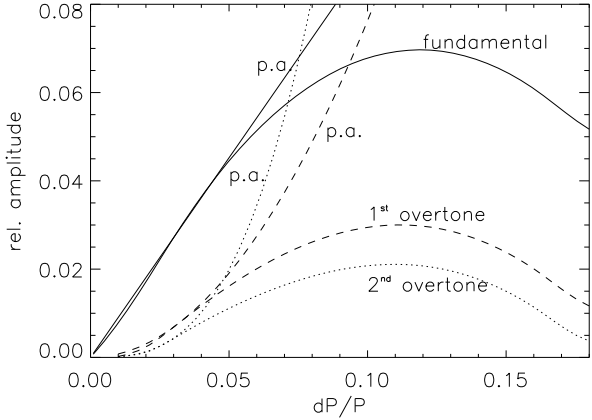}
\caption{Amplitudes of the Fourier coefficients relative to the mean
flux for a local column.  The numerical results are labeled with the
name of the harmonic.  The curves marked with {\sl p.a.} indicate the
linear and quadratic relations predicted by the perturbation
analysis.}
\label{Fig3}
\end{figure}

\begin{figure}[htb]
\includegraphics*[width=8.8cm]{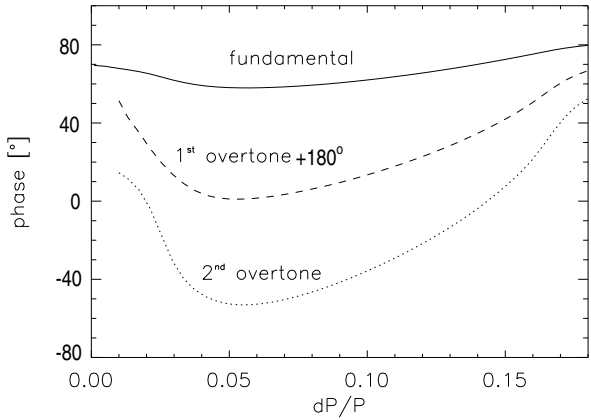}
\caption{ Phases of the fundamental and overtones relative to the
pressure variation in a local column for the numerical results. The
linear perturbation analysis predicts constant phases with values,
which depend on the equilibrium model.}
\label{Fig4}
\end{figure}

\begin{figure}[htb]
\includegraphics*[width=8.8cm]{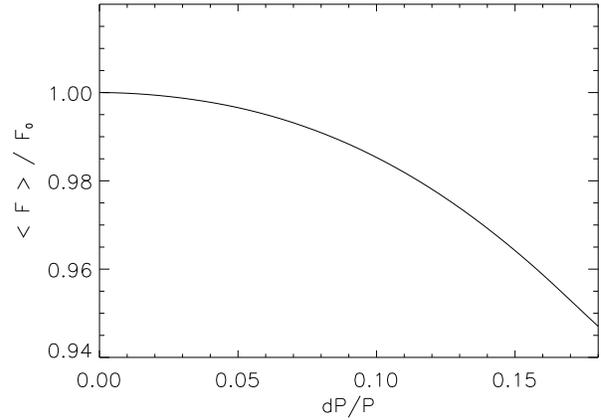}
\caption{
Time--averaged flux relative to the flux below the convection zone.}
\label{Fig5}
\end{figure}

\section{Numerical results for a column}
In this section we study the reaction of the convection zone and the
shape and amplitude of the surface flux variation for a column
on the stellar surface.

For small amplitude pulsations with $1\%$ pressure variation (see
upper panel in Fig.~\ref{Fig1}) the photospheric flux (solid line) is
shifted in phase relative to the flux of the adiabatic layers (dashed
line) and the amplitude is reduced. As shown by Brickhill
(\cite{Brick83}), the phase shift and the reduced amplitude is caused
by the delay in the heat transfer due to the necessary adjustment of
the convective layers to a change in the input flux.  If the flux is
increased, the depth of the convective layer is shrinking as required
by a static envelope solution corresponding to a higher effective
temperature. A large amount of heat is necessary to change
convective into radiative layers and to increase the heat content of
the remaining convection zone, which has to be supplied by the input
flux.  Consequently, the photospheric flux follows the input flux with
a delay of the order of the thermal time scale of the convective
layer. If the input flux varies periodically with a period comparable
to the thermal time scale, this delay results in a phase shift and an
amplitude reduction.

Fig.~\ref{Fig2} summarizes this variation of the light curves with
changing pressure amplitudes.

For small amplitudes the variation of the surface flux remains
sinusoidal and the amplitude increases linearly with the pressure
amplitude as predicted by the perturbation analysis.

For larger amplitudes, e.g. $5\%$ (lower panel Fig.~\ref{Fig1}) the
extent of the the convective zone varies significantly during the
pulsation cycle, and the stratification may even become completely
radiative shortly after maximum compression.  At that instant the
photospheric flux follows directly the adiabatic flux variation,
causing the steep increase and sharp maxima in the light curves.

Increasing the pressure amplitudes even further ($20\%$), the surface
flux variation becomes sinusoidal again, with small amplitudes. In
this case, a large fraction of the heat flux is transformed into
mechanical energy, and the remaining average flux corresponds to a
model with lower (time--averaged) effective temperature and thicker
convection zone. The thermal time constant of this convection
zone  becomes large compared to the pulsation period, resulting in a
strong reduction of the flux amplitude.

We emphasize that this conversion of heat to mechanical energy in the
column (which remains qualitatively the same with the higher numerical
accuracy of the IDE method) does not directly tell us anything about
the global excitation or damping of the mode. Our calculation is based
on energy arguments in an open system: the enforced variations of the
pressure and column cross section carry away any difference of
incoming and outgoing heat flux. In order to really study the
excitation of the mode one would need to calculate the behavior of the
mode throughout the whole star. It is quite possible that the increase
of the mechanical energy in the outer envelope is more than balanced
by a damping in the interior; it is also possible that the energy goes
into the excitation of other modes. This problem will be studied
further in future simulations, which include the study of the $P\,dV$
term in Eq.~\ref{eqn1}.

The sharp maxima in the light curves imply the appearance of higher
harmonics in the Fourier decomposition. Fig.~\ref{Fig3} demonstrates,
how the relative amplitudes of fundamental, first, and second overtone
increase with increasing pressure amplitude. For low amplitudes the
results confirm the predictions of the perturbation analysis (linear
and quadratic dependence for fundamental and first overtone); for
amplitudes larger than $5\%$, however, strong deviations become
apparent.

The sign and even the magnitude of the phase shifts in
Fig.~\ref{Fig4} approximately agree with the predictions of the
perturbation analysis: the fundamental for the flux is lagging behind
the pressure, whereas the in the first harmonic it is
leading. However, quite unexpectedly from the predictions of Goldreich
and Wu, for very small amplitudes the phases do not become constant.
The shift demonstrates the change of the relative contributions from
the compression in the convection zone, which is strictly in phase,
and the flux variation below the convection zone, which is phase
shifted during the propagation through the convective layer. While at
very low amplitudes the numerical result may be questionable, we are
confident that this is not the case for pressure amplitudes larger
than about 1\%. This result needs further study, and we will attempt
in a future paper to reconcile the results of numerical and
perturbational analysis.

Another difference to the perturbation analysis of Gold\-reich and Wu is
the dependence of the mean local flux on the pressure amplitude (see
Fig.~\ref{Fig5}). The flux from below the convection zone is reduced
by a fraction of energy that is converted into mechanical energy to
drive the pulsation mode. This fraction grows quadratically with the
pressure amplitude.  Consequently the mean flux is reduced very
efficiently for large amplitudes.

\section{Surface distribution}
\begin{figure}[htb]
\includegraphics*[width=7.5cm]{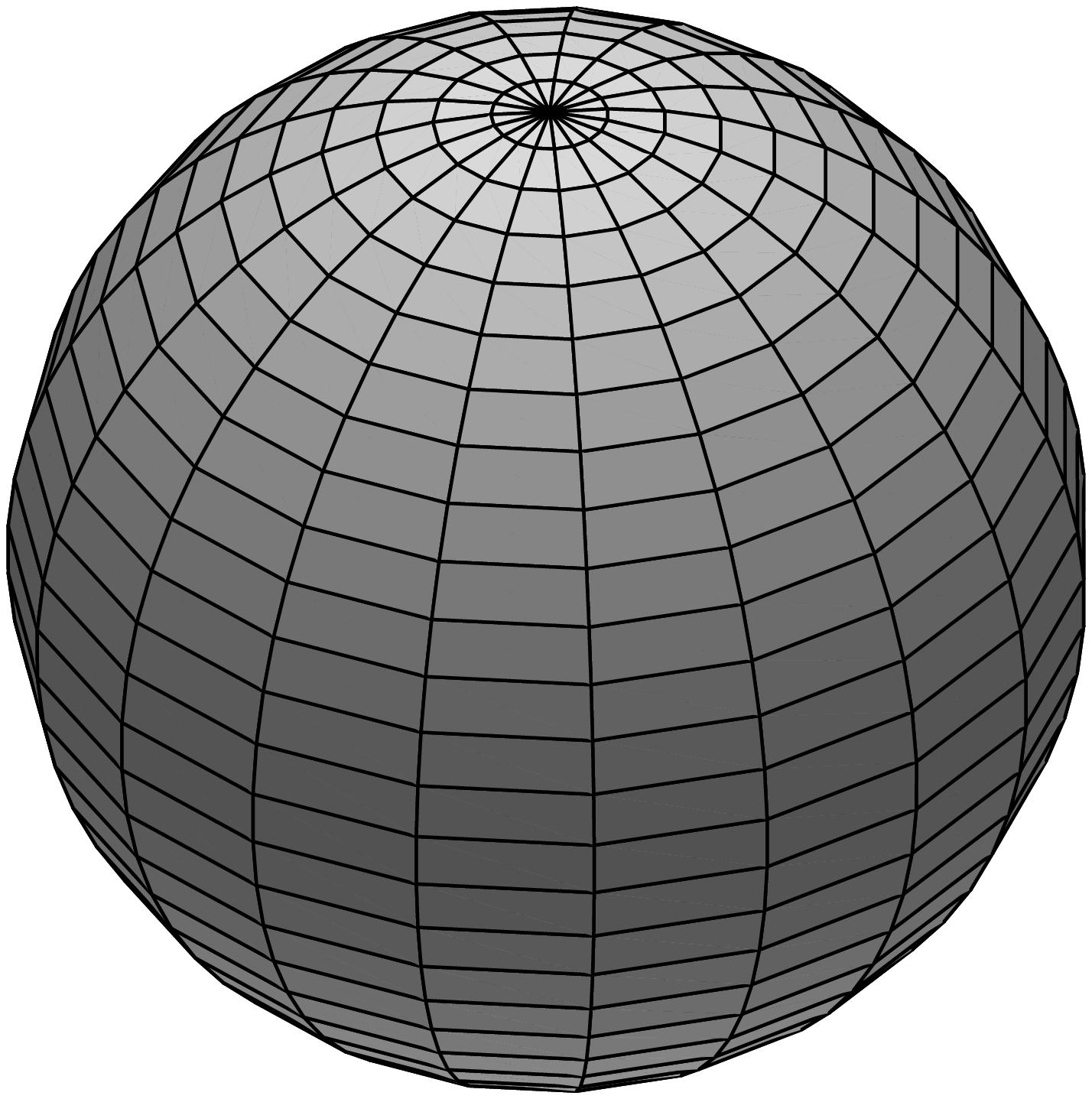}
\includegraphics*[width=7.5cm]{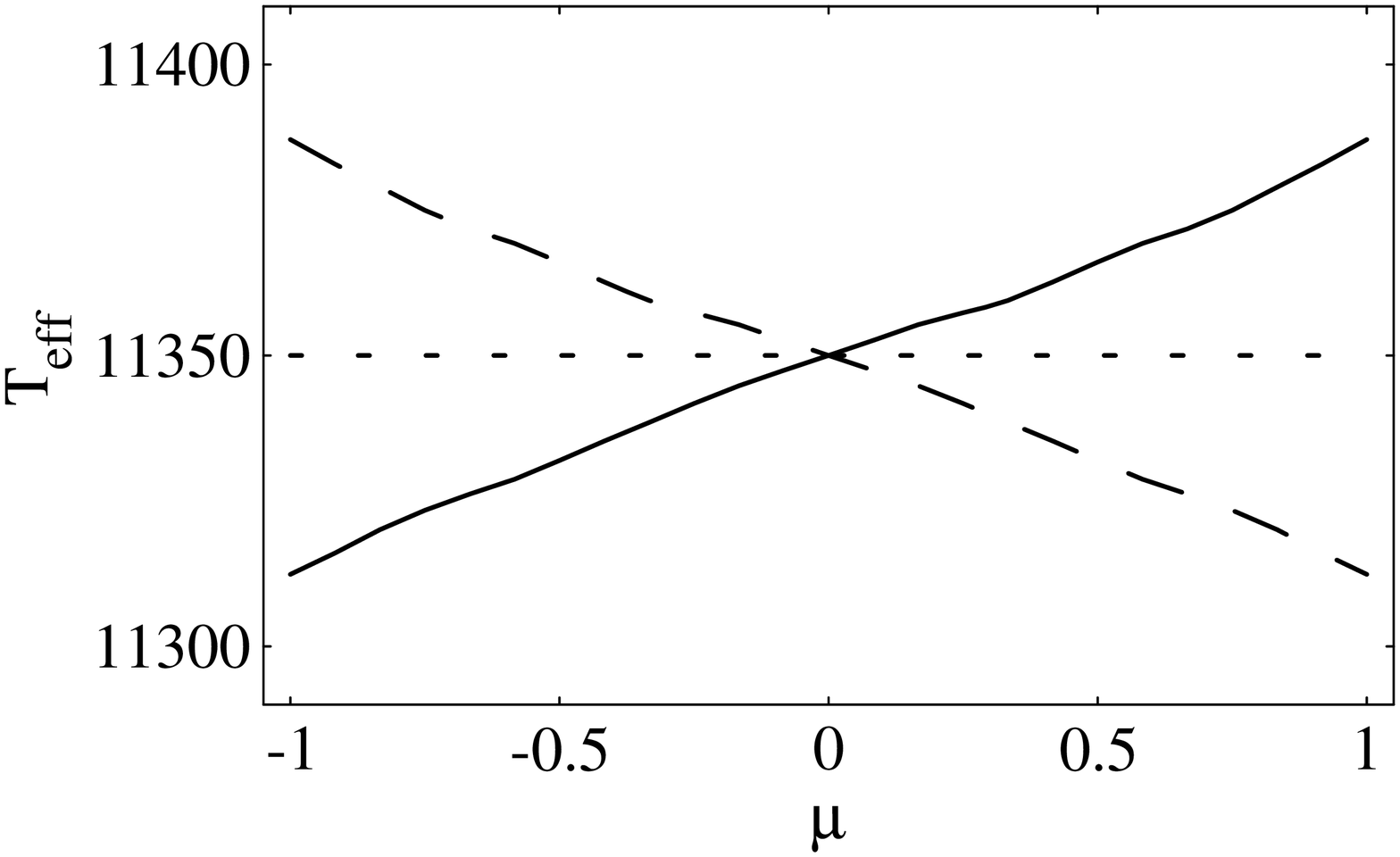}
\includegraphics*[width=7.5cm]{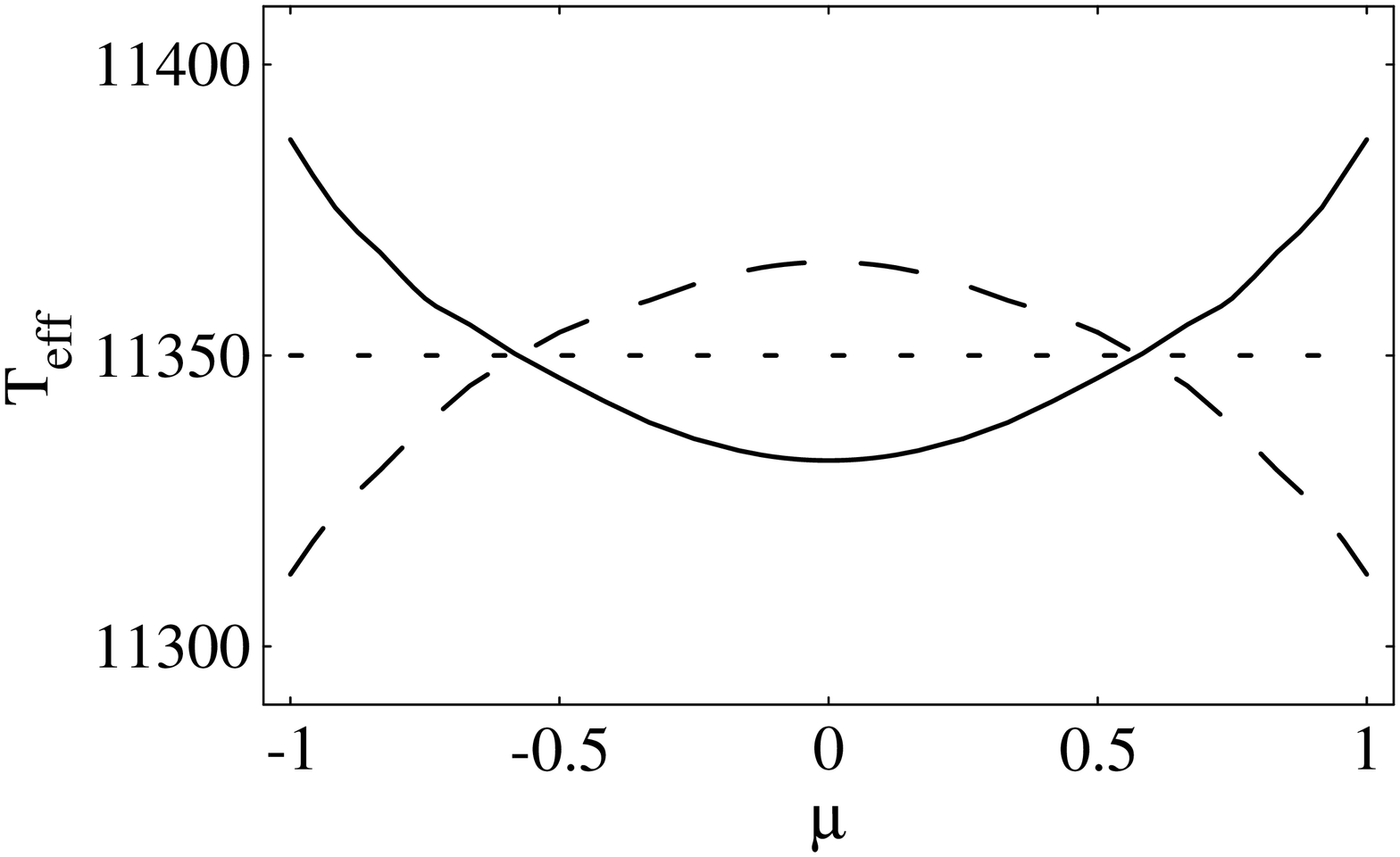}
\caption{ Surface distribution of the effective temperature near the
maximum of the integrated flux. The gray-level of the plot (top, for
an $l=2$ mode) is normalized to the maximum \Teff (white) and minimum
\Teff (black) on the surface. The middle panel shows in graphical form
the temperature distribution with angle $\mu = \cos \vartheta$ for
$l=1$ at three different times, the bottom panel the same but for $l=2$.
The maximum pressure amplitude is 2\% (small amplitude).}
\label{Fig8}
\end{figure}
\begin{figure}[htb]
\includegraphics*[width=7.5cm]{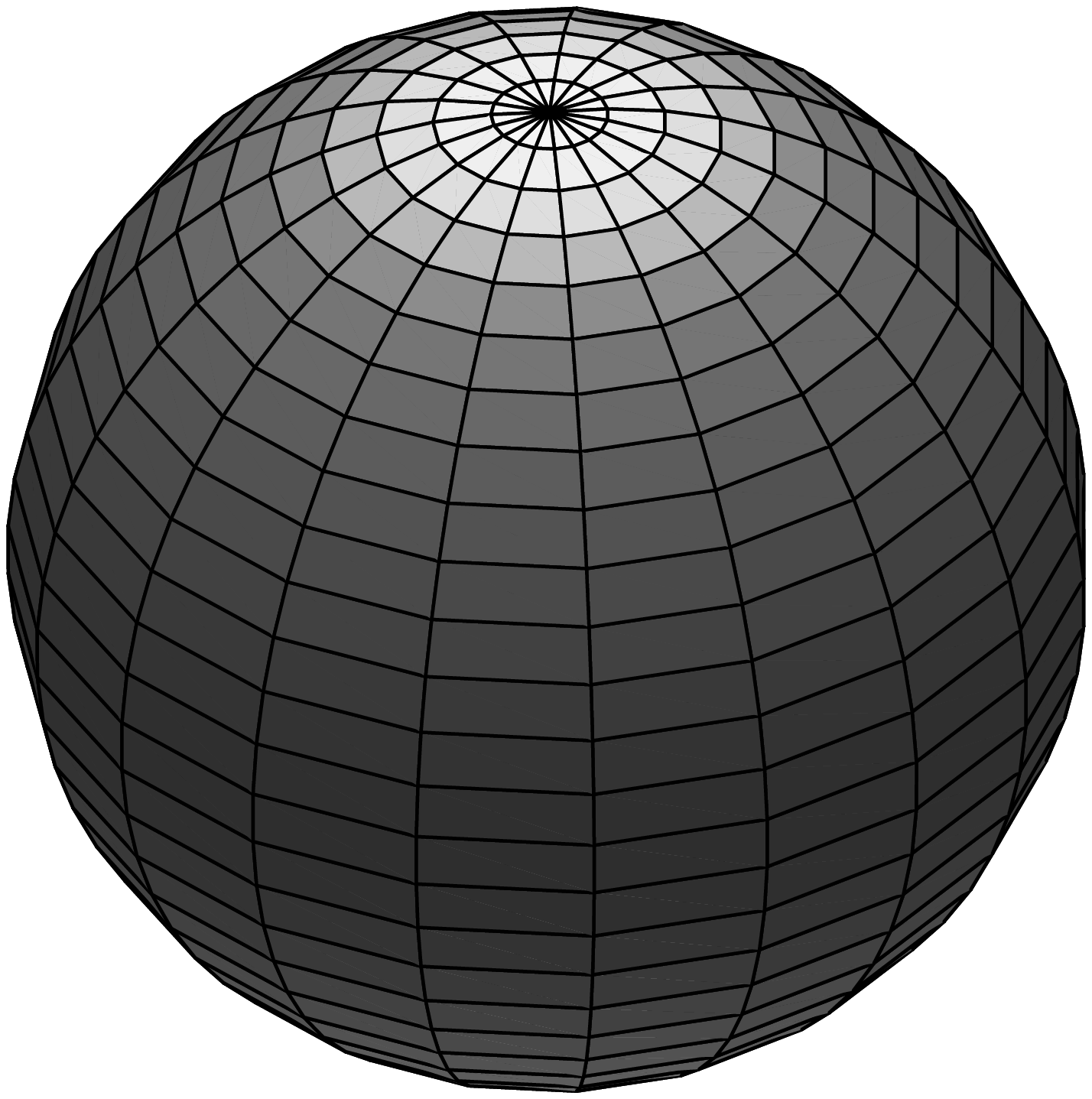}
\includegraphics*[width=7.5cm]{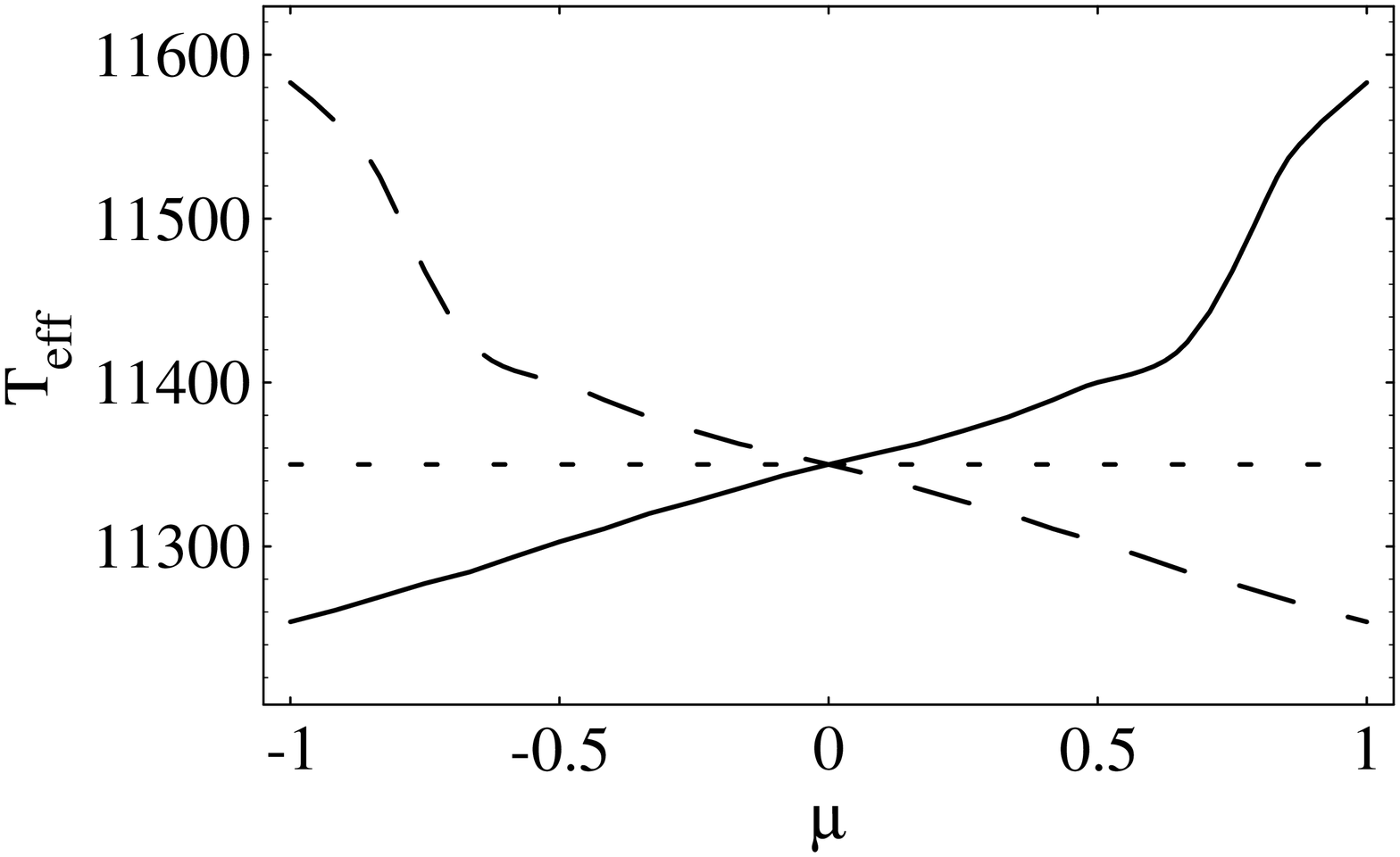}
\includegraphics*[width=7.5cm]{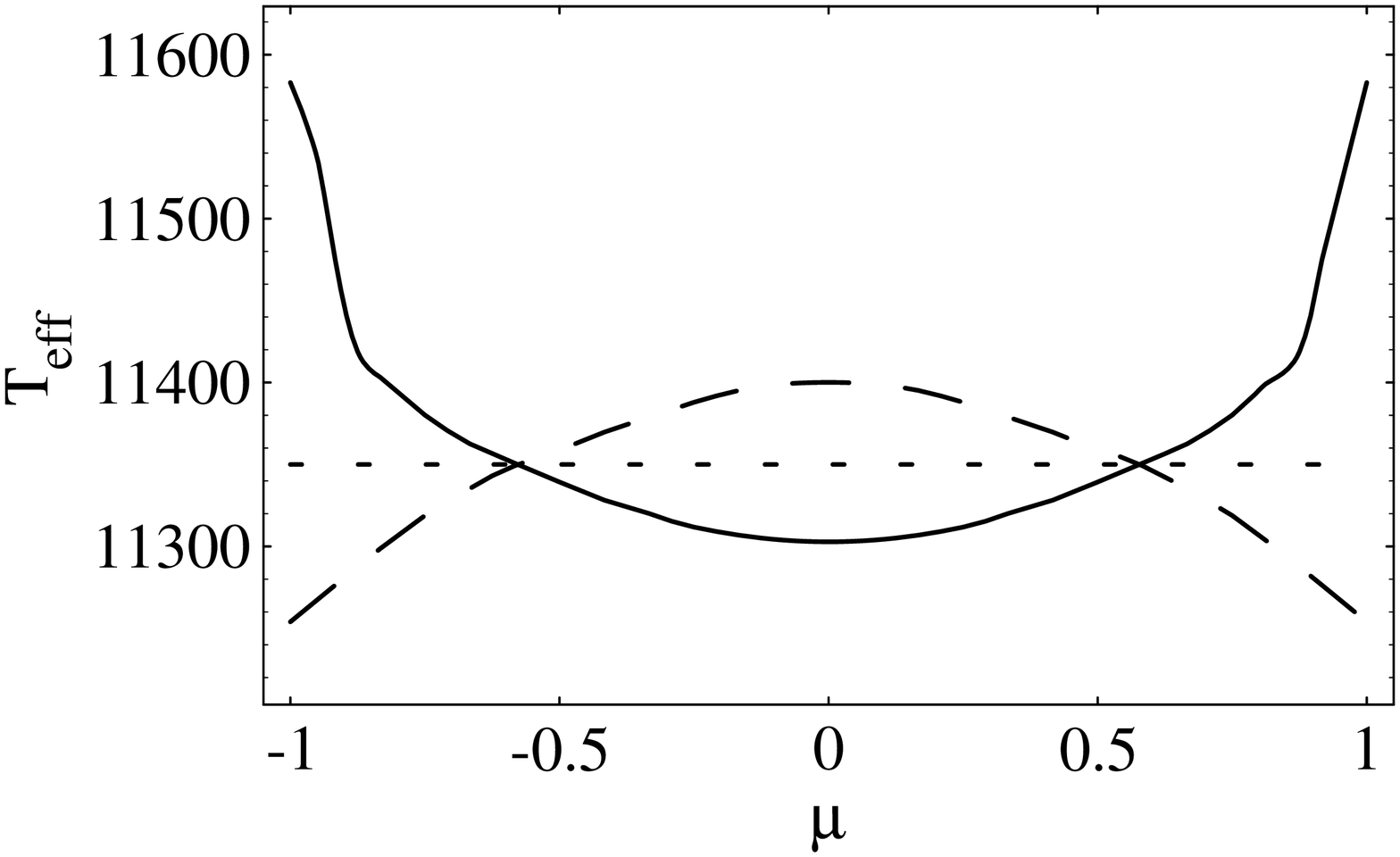}
\caption{Similar to Fig.~\ref{Fig8}, but for a maximum pressure
amplitude of 5\%}
\label{Fig11}
\end{figure}
\begin{figure}[htb]
\includegraphics*[width=7.5cm]{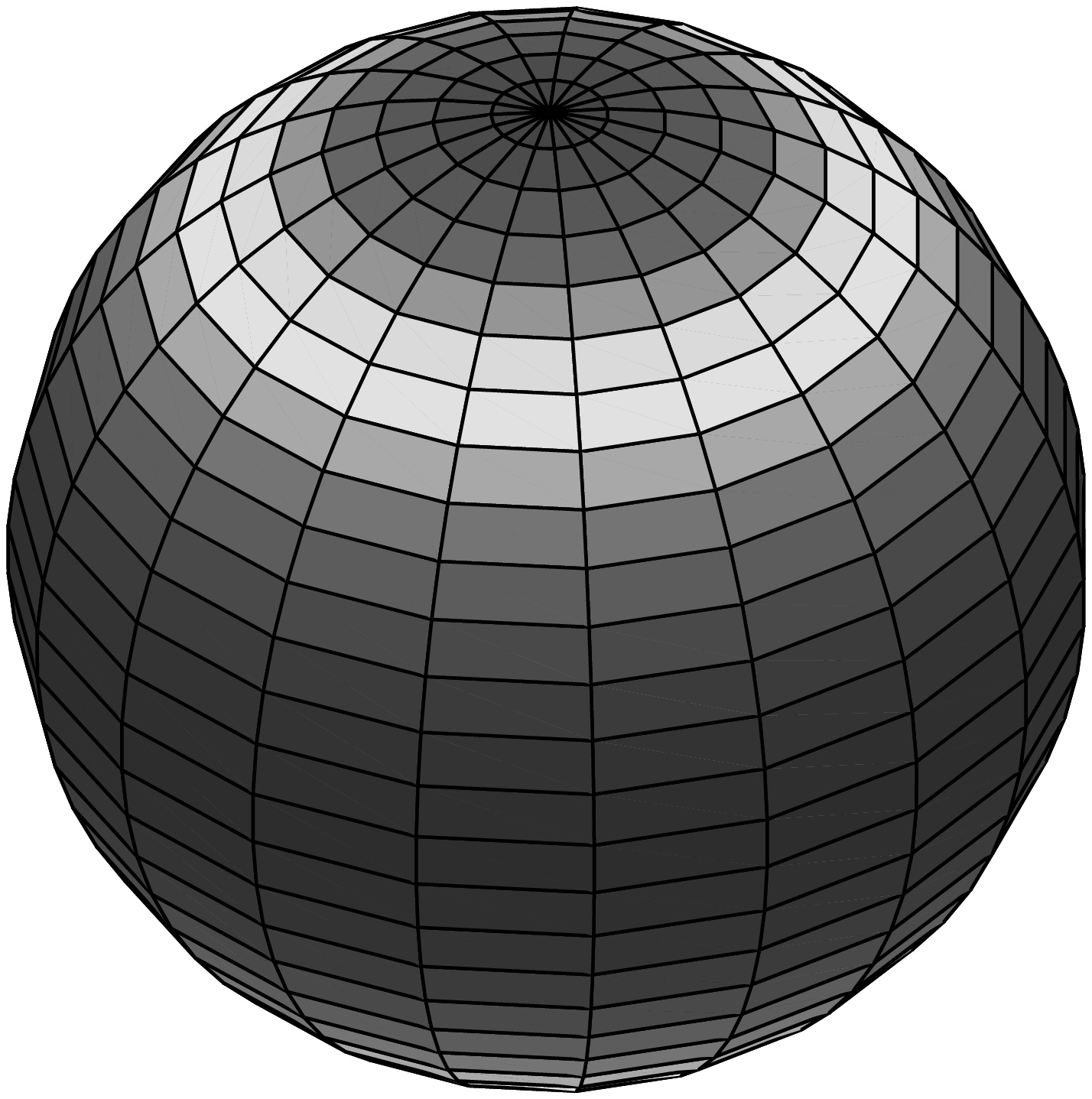}
\includegraphics*[width=7.5cm]{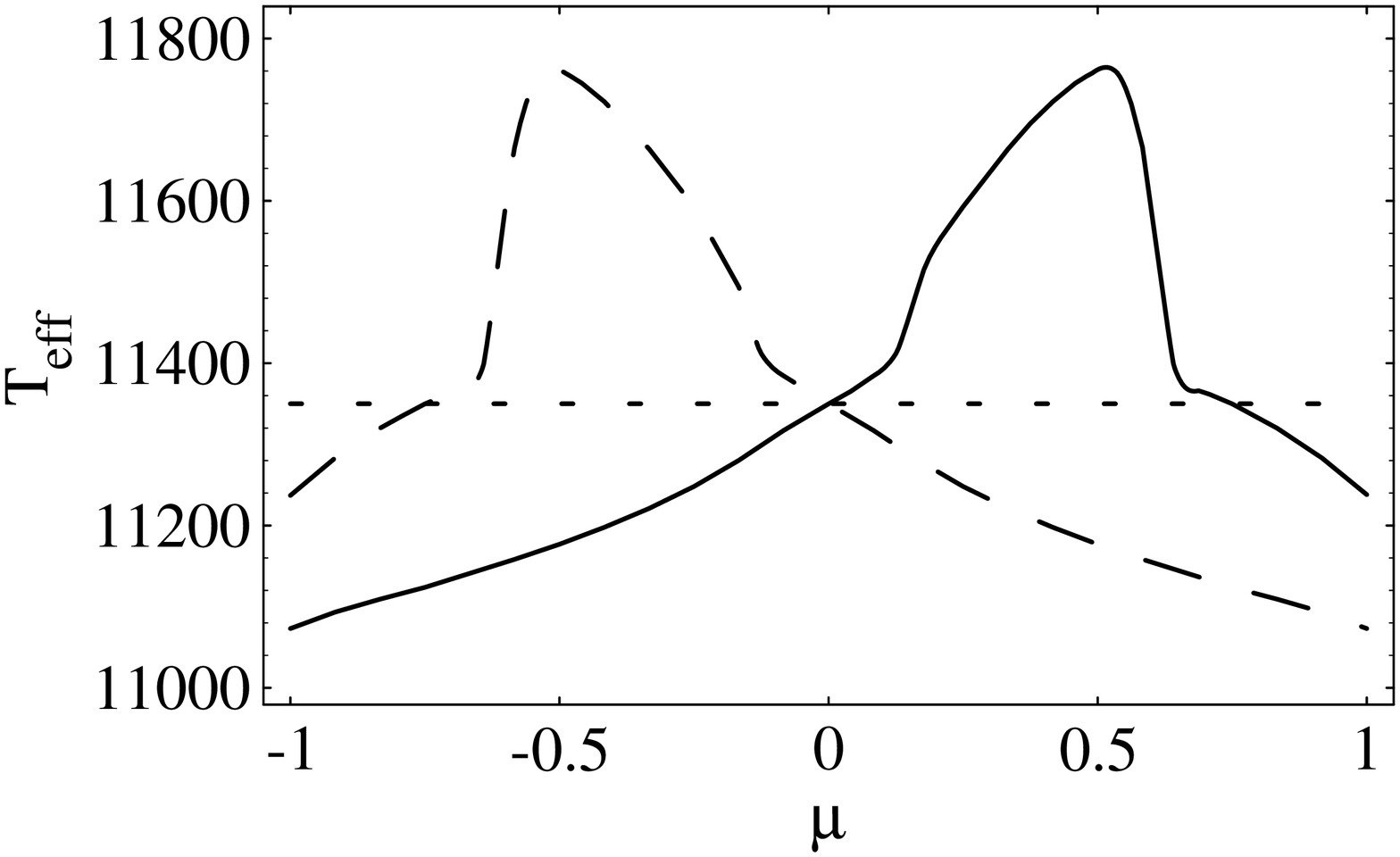}
\includegraphics*[width=7.5cm]{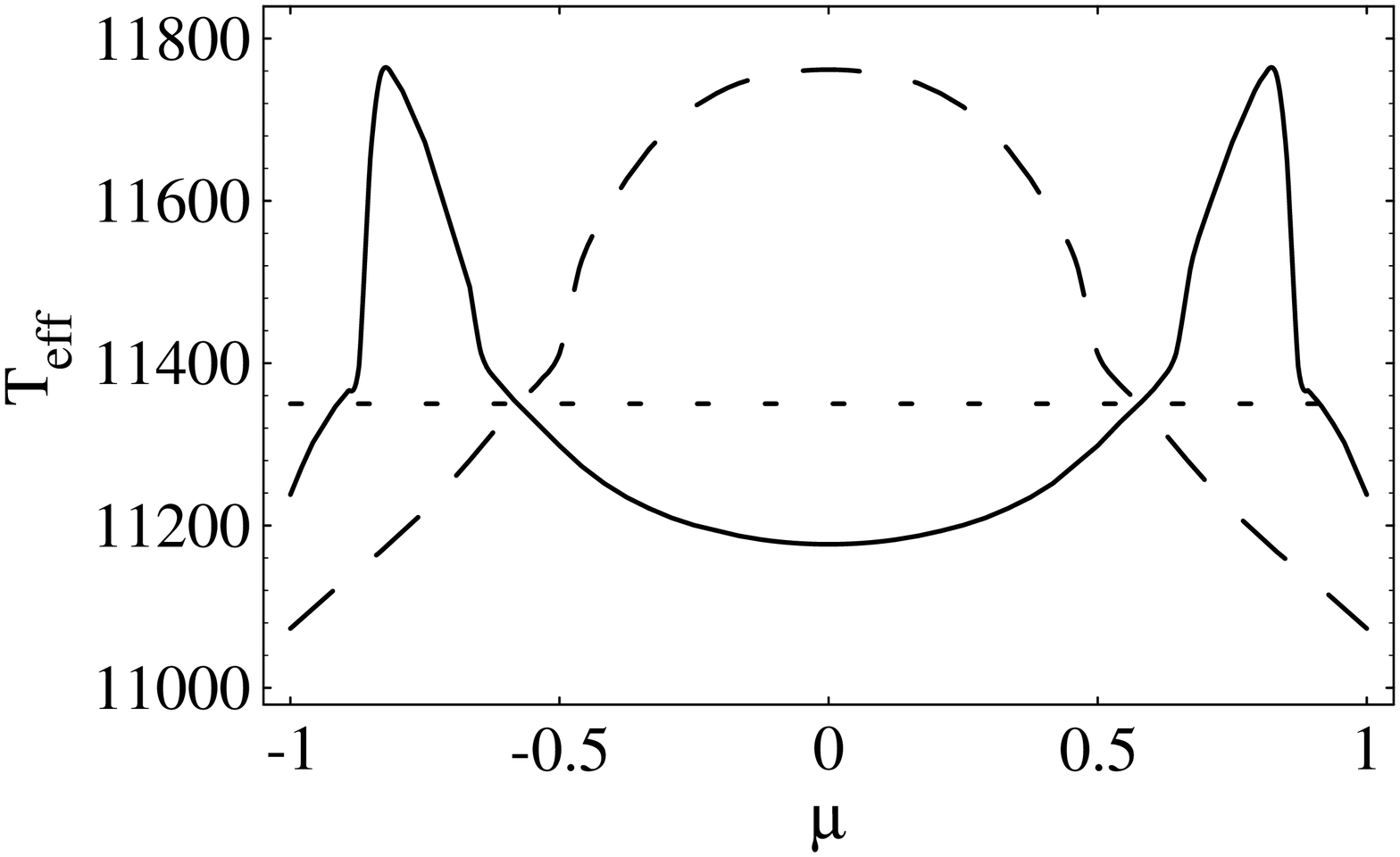}
\caption{Similar to Fig.~\ref{Fig8}, but for a maximum pressure
amplitude of 20\%}
\label{Fig14}
\end{figure}
In the preceding sections we have discussed the local reaction of the
photospheric flux to an assumed fixed pressure amplitude. In this
section we consider the whole surface of the star, assuming that the
spatial structure of the {\em pressure} perturbation as well as the
flux perturbation {\em below} the convection zone is described by a
spherical harmonic of mode numbers $l$ and $m$.

For small amplitudes --- because of the linear dependence of surface
amplitudes on pressure amplitudes --- the distribution of the surface
flux amplitude follows directly the pressure and is therefore
described by the same spheri\-cal harmonic. But this is only true if
one ignores the small phase changes of the flux variation. Due to this
phase shift the maximum flux amplitude is reached at a later time for
smaller amplitudes. This effect leads to a deviation of the
photospheric flux distribution from the spherical harmonic of the
mode, but the flux amplitude at the surface is still a monotonous
function of the pressure amplitude. This changes with increasing
amplitude: because of the non--monotonous local relation the maximum
flux amplitude at the surface does not coincide with the maximum of
the pressure amplitude and e.g. for $m=0$ modes the maximum flux
amplitude is not reached at the poles. Since also the phases change
with amplitude, different latitudes on the surface reach the maximum
flux at different times.

Fig. \ref{Fig8} to \ref{Fig14} show the surface distribution for
$l=1,2, m=0$ modes as three-dimensional plot with gray-level
indicating the effective temperature, and in graphical form the
dependence on polar angle $\mu = \cos \vartheta$. Shown are three
different cases for small, medium, and large pressure amplitudes.

For $l=2$ modes the change due to the nonlinear relation between
pressure amplitude and surface flux amplitude may be quite dramatic:
whereas for small amplitudes the maximum amplitudes are at the poles,
with a smaller relative maximum at the equator, the situation can be
completely reversed at large amplitudes (see Fig.~\ref{Fig14}).

\section{Flux integration and time--resolved spectra}
From these results it follows that the surface distribution of the
flux variation in general has a much more complicated form than is
assumed traditionally (meaning spherical harmonics describing all
variations on the stellar surface). In this section we want to answer
two questions: First we want to find an optimal strategy to calculate
the wavelength dependent amplitude spectra, and second we want to
discuss the major effects on time--resolved spectra one has to expect,
if the linear assumption is not any longer appropriate.

The general problem is to find the time--dependent flux $\flux_\lambda
= \flux_\lambda(\rtg,t)$, where $\rtg$ is the direction to the
observer. To integrate the flux we need the specific intensity at each
point of the surface at each time in the direction $\rtg$.  In
general, the specific intensity is a function of wavelength $\lambda$,
the spatial point $\rvec$, the direction $\rtg$ and the time. We want
to restrict this to a simpler situation, where the intensity can be
calculated from a plane--parallel model atmosphere at each
point. Furthermore we assume, that the surface structure is equivalent
to a static model atmosphere down to optical depths larger than 1 for
each time step. In this case the specific intensity is only a function
of the wavelength, the local flux $F$ and the cosine between $\rtg$
and the vector normal to the surface $\nvec$
\[ I_\lambda = I_\lambda(F,\mu), \]
with $\mu=\nvec\cdot\rtg$. In this approximation the dependence of the
specific intensity on the surface element specified by the angles
$\theta$ and $\phi$ and the time is not explicit but arises from the
the dependence of the local flux on these quantities.

The surface coordinates $(\theta, \phi)$ can be taken as identical to
the coordinates in the spherical harmonic of the pulsation mode. For
the integrated flux it is convenient to introduce a coordinate system
with the axis oriented towards the observer. The system $(\vartheta,
\varphi)$ is defined by
\[ \vartheta = 0 \qquad \mbox{for}\quad \mu = 1. \]
In this case the integrated flux can be written as
\begin{equation} \label{eqnlight}
\flux_\lambda(\rtg,t) = \int^{2\pi}_0 d\varphi \int^1_0 d\cos\vartheta \,
\cos\vartheta \, I_\lambda(F(\vartheta,\varphi,t),\mu).
\end{equation}
Assuming that the flux $\flux_\lambda$ is strictly a periodic function
with the circular frequency $\omega$ we can expand the flux in a
Fourier series
\[ \flux_\lambda(\Omega,t) = \frac{A_0}{2} + \sum_{n=1}^\infty \left[
A_n \cos n \omega t + B_n \sin n \omega t \right]. \]
with 
\begin{equation} \label{eqnA}
A_n = \frac{\omega}{\pi} \int_{t_0}^{t_o+2\pi/\omega}dt\,
\flux_\lambda(\Omega,t)\, \cos n\omega t
\end{equation}
and
\begin{equation} \label{eqnB}
B_n = \frac{\omega}{\pi} \int_{t_0}^{t_o+2\pi/\omega}dt\, 
\flux_\lambda(\Omega,t) \,\sin n\omega t.
\end{equation}
$A_0/2$ is the mean surface-integrated flux.
The amplitude for each harmonic $n$ is defined by
\begin{equation} \label{eqndk1}
\frac{\delta \flux}{\flux}_n \equiv \frac{2 \sqrt{A_n^2 + B_n^2}}{A_0}
\end{equation}
and the normalized amplitude spectrum by
\begin{equation} \label{eqnQ}
\spec_n \equiv \frac{\frac{\delta \flux}{\flux}_n(\lambda)}{
\frac{\delta \flux}{\flux}_n(\lambda_0)} = \frac{A_0(\lambda_0)}{A_0(\lambda)}
\frac{\sqrt{A_n^2(\lambda) + B_n^2(\lambda)}}{
\sqrt{A_n^2(\lambda_0) + B_n^2(\lambda_0)}}.
\end{equation}
Here $\lambda_0$ is an arbitrary reference wavelength, often taken in
the visual at 5500~\AA\ corresponding to the $V$ magnitude.

\subsection{Relation of $I_{\lambda}$ to the local flux $F$}

For the evaluation of Eq. (\ref{eqnlight}) it is necessary to connect
the specific intensity with the total flux of a local column. 
Using an expansion to first order we assume for this relation
\begin{equation} \label{eqna1}
I_\lambda = I_\lambda(F_0,\mu) - \dIdF\!(\mu)\,F_0 + \dIdF(\mu)\,F(t)
\end{equation}
Expanding the local flux in a Fourier series in the same manner as for
the surface integrated flux
\[ F(t) = \frac{a_0}{2} + \sum_{n=1}^\infty \left[
a_n \cos n \omega t + b_n \sin n \omega t \right] \]
the amplitude for the harmonic $n$ is
\begin{equation} \label{eqnlf}
\frac{\delta F}{F}_n \equiv \frac{2 \sqrt{a_n^2 + b_n^2}}{a_0}.
\end{equation}

Inserting (\ref{eqna1}) into (\ref{eqnlight}) one obtains after some
manipulation (Wu \cite{Wu98}) for $n > 0$ the very simple result 
\begin{equation} \label{eqnl2a}
A_n = \int_0^{2\pi}d\varphi \int_0^1 d\cos\vartheta \, \cos\vartheta 
\dIdF\!(\mu(\vartheta,\varphi))\, a_n
\end{equation}

\begin{equation} \label{eqnl2b}
B_n = \int_0^{2\pi}d\varphi \int_0^1 d\cos\vartheta \, \cos\vartheta
\dIdF\!(\mu(\vartheta,\varphi))\, b_n.
\end{equation}

The integration for $A_0$ leads to two additional terms and reduces to
the simple form above only if the local time--averaged flux $a_0/2$ is equal
to the equilibrium flux $F_0$ for the local column.

Using  approximations similar to those of Robinson et
al. (\cite{Robi82}) we could simplify the relation for the intensity as
\begin{equation}\label{eqnla1}
 I_\lambda = h_\lambda(\mu) \, F
\end{equation}
with a limb darkening function $h_\lambda$. The major simplification
is that in this form the limb darkening does not depend on the total
flux and the intensity derivative in Eq.(\ref{eqna1}) is
simply given by $h_\lambda$. In this case the integral for the global
coefficient $A_0$ always reduces to the simple form of
Eq.(\ref{eqnl2a}). However, we will show below that this approximation
is not adequate, since the the amplitude spectra depend very
sensitively on the limb darkening and in the derivative of
Eq.(\ref{eqnla1})
\begin{equation}
\dIdF = h_\lambda(\mu, F_0) + \left.\frac{\partial h_\lambda}{\partial
F}\right|_{F_0} 
\end{equation}
the second term is in general not negligible.

\subsection{The linear assumption}
The linear assumption is to specify the local flux by
\begin{equation} \label{eqna3}
\frac{\delta F}{F} =  \frac{\delta F}{F}_1^{\rm max} Y_l^m(\theta,\phi) 
\, e^{i\omega t}
\end{equation}

If we want to determine the global flux amplitudes directly from the
local amplitudes without the intermediate step of Fourier coefficients
we need to make further assumptions. If the coefficient $a_0$ and the
ratio $a_n/b_n$ are constant over the stellar surface, it is possible
to change the order of the square root and the integration over the
visible disk in Eqs.(\ref{eqndk1},\ref{eqnl2a},\ref{eqnl2b}). After
some algebraic manipulations we get
\begin{equation} \label{eqna2b}
\frac{\delta \flux}{\flux}_n = \frac{
\int_0^{2\pi} d\varphi \int_0^1 d\cos\vartheta\,
\cos\vartheta \dIdF\!(\mu(\vartheta,\varphi)) \,\frac{\delta F}{F}_n}{
\int_0^{2\pi} d\varphi \int_0^1 d\cos\vartheta\,
\cos\vartheta \dIdF\!(\mu(\vartheta,\varphi))}.
\end{equation}

\begin{figure}[htb]
\includegraphics*[width=8.8cm]{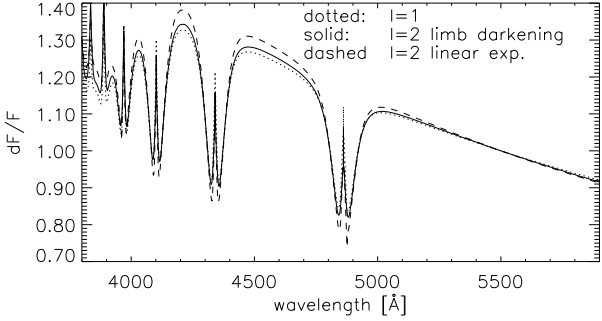}
\caption{Comparison of the amplitude spectra for an $l=2$~mode,
calculated with a limb darkening independent of \Teff (solid line)
and with the linear expansion of the intensity (dashed line). The spectra
for the $l=1$~mode are very similar and represented by the dotted line
in the plot.}
\label{Fig15}
\end{figure}
It is common to transform the spherical harmonic $Y_l^m(\theta,\phi)$
to a system of spherical harmonics in the coordinate system
$(\vartheta,\varphi)$ by
\[ Y_l^m\!(\theta,\phi) = \sum_{-l \le m' \le l} R^l_{m' m} 
Y_l^{m'}\! (\vartheta,\varphi) \] As was shown in several papers
(e.g. Robinson et al. \cite{Robi82}) all terms with $m' \neq 0$ do not
contribute to a flux variation in the integrated light, because they
are equivalent to a rotating pattern with respect to an axis directed
to the observer. For the term with $m'=0$ the assumption of a constant
phase over the stellar surface is valid and the Fourier coefficient
$a_0$ is assumed to be constant over the surface by Eq.(\ref{eqna3})
implicitly. Consequently, we can use Eq.(\ref{eqna2b}) to calculate
the amplitude of the integrated flux with the local amplitude
\begin{equation} \label{eqna4}
\frac{\delta F}{F}'_1 = R^l_{0 m}\, \frac{\delta F}{F}_1^{\rm max}
N_l \LegP_l(\cos\vartheta).
\end{equation}
The quantity $N_l$ symbolizes the normalization of the spherical
harmonic $Y_l^0$
and $\LegP_l$ is the Legendre polynomial.
The only inclination dependent term in Eq.(\ref{eqna4}) is the
coefficient $R^l_{0 m}$. The amplitude $(\delta F / F)_1^{\rm max}$
and $R^l_{0 m}$ are constant over the whole surface and wavelength
independent. Consequently, the normalized spectra $\spec$ are
independent of amplitude and inclination. Fig.\ref{Fig15} shows one
example of the results for a flux independent limb darkening and for
the more general form of the linearized intensity for the model
parameters we use in this paper, but with the linear assumption of
Eq.(\ref{eqna3}). Both results are identical for an $l=1$~mode, but
differ for larger $l$. Even for the $l=2$~modes, there are significant
differences.

To test the accuracy of the linear approximation for the intensity
(Eq.(\ref{eqna1})) we have calculated the
Fourier coefficients directly from the integrated light curve by means
of Eq.(\ref{eqnlight}), (\ref{eqnA}) and (\ref{eqnB}) and have
compared the result with the results from the linearized intensity. We
find a very good correspondence for all wavelengths for modes with $l
\le 3$ and amplitudes less than $30\%$. This range is much larger
than the range, where the total flux variation can be assumed to be
linear with \Teff. For even larger amplitudes the amplitude spectrum
calculated directly from Eq.(\ref{eqnlight}) becomes inclination and
amplitude dependent, because the function $h_\lambda$ varies over the
surface for different flux amplitudes. Such large amplitudes are not
observed, however.

\subsection{The Goldreich and Wu theory}
The Goldreich and Wu theory assumes --- following Brickhill --- a
pressure variation with a spatial and time--dependence described by the
spherical harmonic and period of the pulsation mode. They predict the 
appearance of non--sinusoidal flux variations at the surface 
\begin{equation} \label{eqna5}
\frac{\delta F}{F} = \frac{a_0}{2} + \frac{\delta F}{F}_1 \sin(\omega t - 
\alpha_1) + \frac{\delta F}{F}_2 \sin(2\omega t - \alpha_2) + ...
\end{equation}
An important result of their calculation is that the amplitude of the
fundamental is proportional to the pressure amplitude and the spatial
distribution therefore still given by the same spherical harmonic. The
first overtone varies quadratically with the pressure, which leads to
the appearance of a limited number of spherical harmonics. The phases
$\alpha_n$ are independent of the amplitude of the variation. 

The result for the fundamental is equivalent to the linear assumption
Eq.(\ref{eqna3}) and as a consequence leads to the same results for
the normalized spectra: they depend only on the mode number $l$.

The amplitude of the first overtone has a different behavior than the
spherical harmonic of the mode over the stellar surface. It can be written
as
\begin{equation} \label{eqna7}
\frac{\delta F}{F}_2 = \frac{\delta F}{F}_2^{\rm max} \left[
N_l P_l(\cos\theta) e^{i\alpha_2} \right]^2 \equiv 
\frac{\delta F}{F}_2^{\rm max} V(\theta,\phi)
\end{equation}
The angle dependent function $V$ can be expanded in a series of spherical
harmonics
\begin{equation} \label{eqna8}
V(\theta,\phi) = \sum_{l'=0}^\infty \sum_{m'=-l'}^{l'} c_{l'}^{m'} 
Y_{l'}^{m'}(\theta,\phi).
\end{equation}
It is obvious that in this case $V$ is a polynomial of the order
$2l$ and consequently all coefficients $c_{l'}^{m'}$ vanish for $l' >
2l$, but for this discussion we only need, that the sum has in general
more than one term. Analogous to the linear assumption we can
transform each term of Eq.(\ref{eqna8}) to the coordinate system of
the observer
\begin{equation} \label{eqna9}
V(\theta,\phi) = \sum_{l=0}^\infty \sum_{m'=-l'}^{l'} \sum_{m''=-l'}^{l'}
c_{l'}^{m'} \, R_{m'' m'}^{l'} \,Y_{l'}^{m''}(\vartheta,\varphi)
\end{equation}
The relation Eq.(\ref{eqna9}) can be introduced in Eq.(\ref{eqna7})
and we get an expression for $(\delta F/F)_2$ in Eq.(\ref{eqna2b}).
Again, only the terms with $m'' = 0$ lead to a non--vanishing amplitude
in the integrated light. Inclination dependent quantities are only the
coefficients $R_{0 m'}^{l'}$, but they are different for all $l'$.
Consequently, the inclination dependence does not cancel in the
normalized spectrum $\spec_2$ and the spectrum becomes inclination
dependent. In contrast to this, the amplitude $(\delta F/F)_2^{\rm
max}$ cancels out, and the normalized amplitude spectrum remains
independent of the amplitude.

The 2$^{\rm nd}$ overtone will not be discussed in detail. The local
flux amplitude $(\delta F/F)_3$ cannot be represented by a single
$(\delta F/F)^{\rm max}$ and consequently the normalized amplitude
spectra become in general also amplitude dependent.

\section{Amplitude spectra from numerical calculations}
To determine the normalized amplitude spectra from our numerical
results, it is possible to use the integration of the local Fourier
coefficients $a_n$ and $b_n$.  However, because the phase is not
constant (see Fig.~\ref{Fig4}) over the surface and $a_0$ is a
function of the coordinates $(\theta,\phi)$ (see Fig.~\ref{Fig5}) the
expression Eq.(\ref{eqna2b}) is completely useless even for
$m=0$~modes and we have to calculate the coefficients $A_n$ and $B_n$
separately.  This means that there is no economical advantage to
calculate the Fourier series locally and then to integrate the
coefficients over the visible disk.

We therefore decided to calculate the flux $\flux_\lambda$ for each
time step completely numerically by using Eq.~(\ref{eqnlight}). This
method also avoids the disadvantage of having to use a linearization
like Eq.~(\ref{eqna1}) to reach expressions for the Fourier
coefficients of the integrated flux $\flux_\lambda$.
From the time--dependent flux we can calculate the Fourier coefficient
with the aid of Eq.~(\ref{eqnA}) and Eq.~(\ref{eqnB}). The numerical
results presented in the following section are based on such
integrations.

On the other hand, in order to gain more insight into the qualitative
behavior of the spectra for different amplitudes and inclinations, we
will use simple fits for the numerical results of the local variation,
which allow then analytic integrations for the total flux.
Both approaches are complementary: with the completely numerical
simulation we avoid simplifications, which may not be correct, but we
can calculate the result only for very few combinations of
parameters. The semi-analytical study, on the other hand, shows the
functional dependence of the results on quantum numbers, inclination,
etc. 

\subsection{General effects on the spectra}
The phase dependence of all local Fourier coefficients and the
nonlinear dependence on the pressure variation leads to deviations
from the assumptions made traditionally for the flux variation over
the stellar surface.  The intention of this subsection is to show, how
this deviation affects the synthetic spectra for different
inclinations, pulsation amplitudes and mode numbers $l$.
\begin{figure}[htb]
\includegraphics*[width=8.8cm]{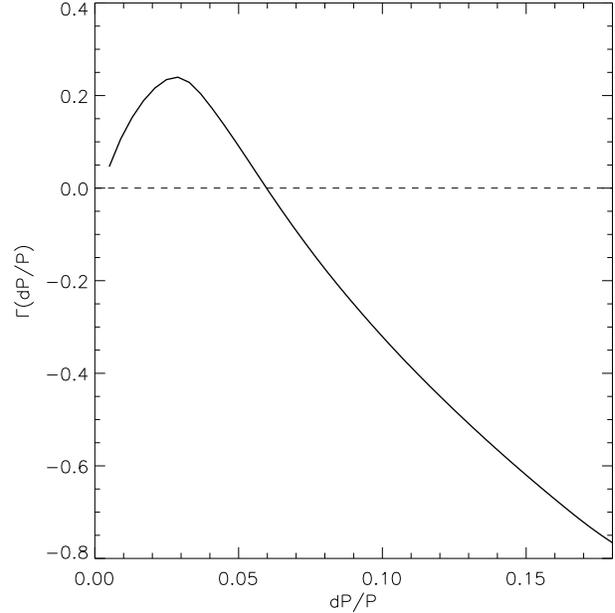}
\caption{ The amplitude dependence of the correction term
$\spec_n^{\rm corr}$. The factor $\Gamma$ is displayed for our model
parameters.}
\label{Fig16}
\end{figure}

To find an expression for the error made when using the linear
assumption, we assume a flux independent limb darkening function and
the integrated Fourier coefficients given by Eqs. (\ref{eqnl2a}) and
(\ref{eqnl2b}). We define a correction term for the fundamental as
\begin{equation} \label{eqndk2}
\spec_n^{\rm corr} \equiv \frac{\frac{\delta \flux}{\flux}_n(\lambda)}{
\frac{\delta \flux}{\flux}_n(\lambda_0)} 
-
\left.\frac{\frac{\delta \flux}{\flux}_n(\lambda)}{
\frac{\delta \flux}{\flux}_n(\lambda_0)} \right|_{\rm linear}.
\end{equation}
In appendix A we demonstrate that in a first order approximation
$\spec_1^{\rm corr}$ can be separated into three factors
\begin{equation} \label{eqna19}
\spec^{\rm corr}_1 = \frac{A_0(\lambda_0)\,\I_0(\lambda)}
{A_0(\lambda)}\,\Gamma\!\left(\p_{\rm max}\right) \,
\Lambda(i,l).
\end{equation}

The first term is practically independent of wavelength ($\I_0$ is
defined in Eq.~\ref{defI} in the Appendix).
The second term $\Gamma$ is derived from the expansion coefficients of
the local flux as a function of the pressure variation. For finite
amplitudes we interpret these coefficients as fit parameters to
describe the local flux as a polynomial of $2^{\rm nd}$ order up to
the maximum pressure variation, that occurs at the stellar surface.
This is plotted in Fig.~(\ref{Fig16}). $\Gamma$ is not a function of
the inclination or the mode numbers, but depends on the stellar
parameter as \Teff\ and the pulsation period. For our model parameters
we can divide the synthetic spectra in three different pressure
regimes: Small amplitudes up to $\delta P/P_{\rm max}\sim 5\%$, where
the function $\Gamma$ is positive; intermediate amplitudes $5\% <
\delta P/P_{\rm max} < 7\%$, where $\Gamma$ is approximately $0$, and
large amplitudes $\delta P/P_{\rm max} > 7\%$ with a negative
$\Gamma$. In the following 3 subsections these cases are discussed
separately for the numerical results.
\begin{figure}[htb]
\includegraphics*[width=8.8cm]{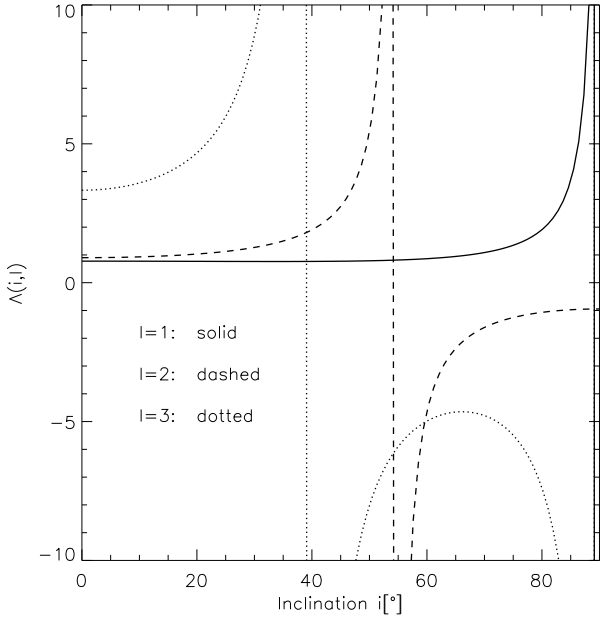}
\caption{ The factor $\Lambda$ is shows the dependence of the
correction factor between the linear assumption and the numerical
simulation on inclination. The 3 curves show the factor for 3
different mode numbers ($l=1,2,3$) and the Edington limb darkening law.
Large deviations occur near an inclination of 55 degrees. 
}
\label{Fig17}
\end{figure}

The third term $\Lambda$ is the inclination dependent term. Besides
this it depends on the mode numbers and wavelength, but not on the
amplitude. The results for the factor $\Lambda$ for $l \le 3$
are plotted in Fig.~(\ref{Fig17}). This factor is for most
inclinations very small for $l=1$ and moderate for $l=2$, if the
inclination is clearly different from the latitude of the
node lines. For $l=3$ the factor $\Lambda$ is much larger and leads to
a large global deviation of the numerical results from those using the
linear assumptions for any inclination and amplitude.

\begin{figure}[htb]
\includegraphics*[width=8.8cm]{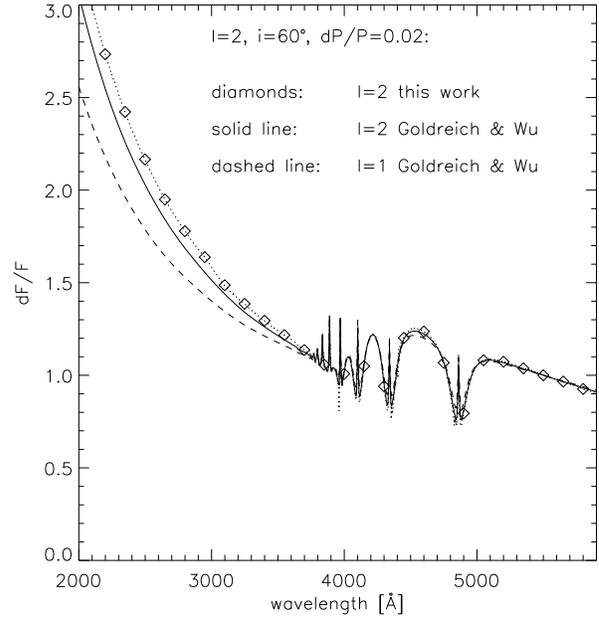}
\caption{Amplitude spectra for small pressure amplitude. The solid
line and the dashed lines are results based on the spherical harmonics
$l=2$ and $l=1$ respectively. The diamonds connected with the dotted
line are the numerical results.}
\label{Fig18}
\end{figure}

\begin{figure}[htb]
\includegraphics*[width=8.8cm]{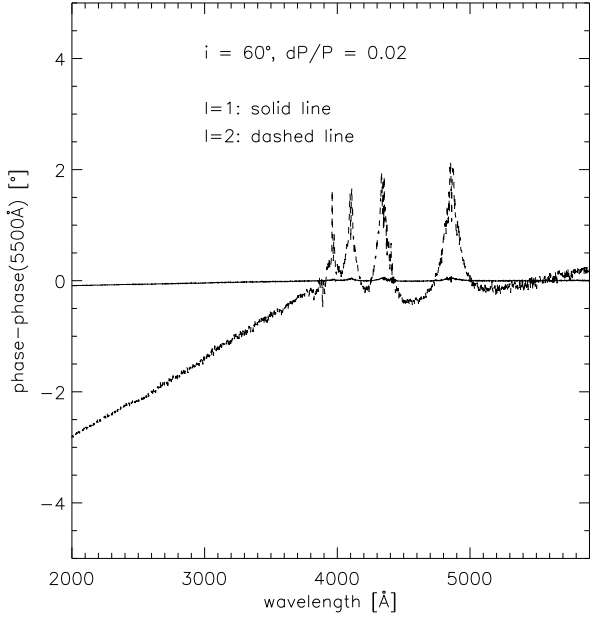}
\caption{Phase spectra for small pressure amplitude. The solid line
($l=1$) and the dashed line ($l=2$) gives the predicted phase shift
for the wavelength range from $2000$\AA to $6000$\AA.}
\label{Fig19}
\end{figure}

\begin{figure}[htb]
\includegraphics*[width=8.8cm]{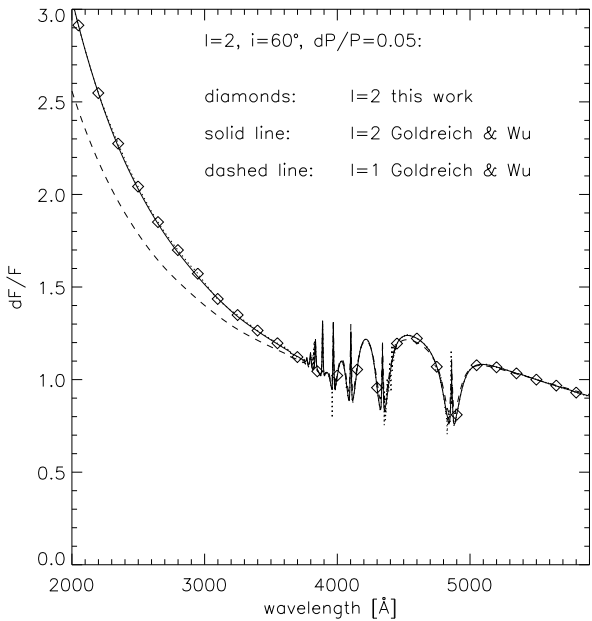}
\caption{Amplitude spectra for intermediate amplitude. The lines have
the same meaning as in Fig~(\ref{Fig18}).}
\label{Fig20}
\end{figure}

\subsubsection{Small amplitudes}
In the previous section we have discussed the effects of the
nonlinear behavior of the light curves on the amplitude spectra of
the fundamental with some approximations. To find the correct spectra
we now calculate the Fourier coefficients directly from the integrated
light curves in a strictly numerical way.

As an illustrating example we take the $l=2$ mode with a maximum
pressure amplitude of $2\%$. For this amplitude the light curves are
sinusoidal in a good approximation, so we can restrict the discussion
to the fundamental of the mode.

From Fig.~(\ref{Fig17}) we do not expect very large deviations from
the linear result for most inclinations. To show the effect we
take an inclination of $60\Deg$ near the latitude of the node line,
where we expect a moderate deviation. The numerical result is plotted
in Fig.~(\ref{Fig18}), together with the result of the linear
assumption, that is supported by the theory of Goldreich and Wu.
All amplitudes in this and the following similar figures have been
normalized to 1 at 5500~\AA.
Although the effect is not very large, it is important for mode
identification, because of the generally small differences between the
calculations for different mode numbers.

Deviations in the spectra already for small amplitudes are suggested
by the analytical calculations in the previous subsection, but are not
obvious in the plot of the absolute amplitude Fig.~(\ref{Fig3}), which
grows linearly with the pressure amplitudes up to $\sim
5\%$ pressure amplitude. Consequently, the deviation has to be an
effect of the phase shift for small amplitudes. This phase shift
should be visible in the integrated light as well. Fig.~(\ref{Fig19})
gives the numerical prediction for the phase shift for the same mode
and modes with $l=1$. In contrast to the prediction of Goldreich and
Wu, we expect a small phase shift of a few degrees over the spectral
range for the $l=2$ modes.

\subsubsection{Intermediate amplitudes}
The amplitude dependent factor $\Gamma$ is very small in this region
and we expect practically no differences to the predictions of
Goldreich and Wu.  Fig.~(\ref{Fig20}) shows the numerical result for
the same mode and inclination as Fig.~(\ref{Fig18}), but for a
pressure amplitude of $5\%$. The spectra for the numerical prediction
and the Goldreich and Wu theory are very similar.

\begin{figure}[htb]
\includegraphics*[width=8.8cm]{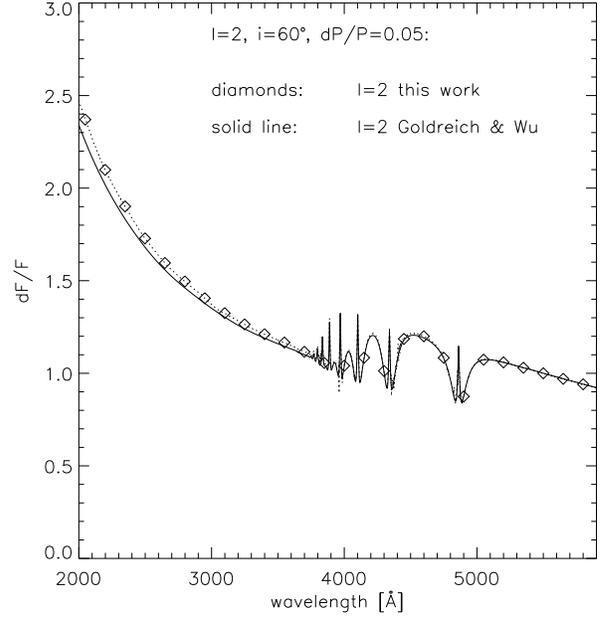}
\caption{Amplitude spectra for the first overtone and intermediate
amplitude.  The lines have the same meaning as in
Fig~(\ref{Fig18}).}
\label{Fig21}
\end{figure}
\begin{figure}[htb]
\includegraphics*[width=8.8cm]{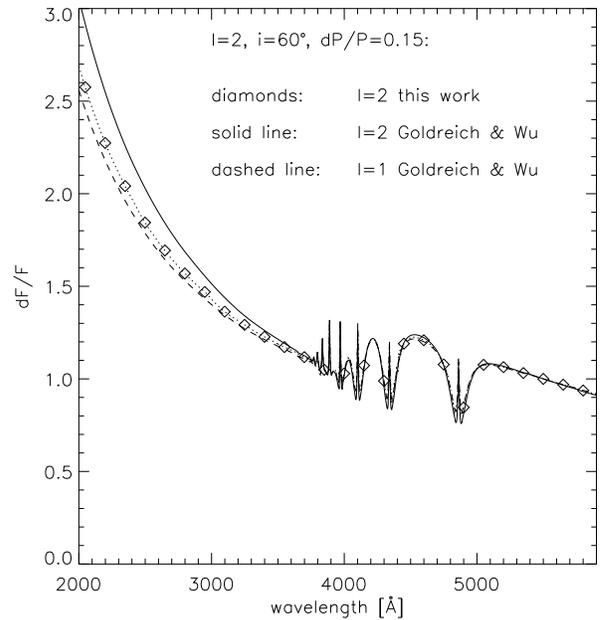}
\caption{Amplitude spectra for large amplitude.
The lines have the same meaning as in Fig~(\ref{Fig18}).}
\label{Fig22}
\end{figure}

In contrast to the small amplitude case, the light curves for $5\%$
pressure amplitude are non--sinusoidal with large flux amplitudes. We
can calculate the expected deviation from the predictions of Goldreich
and Wu analogous to Eq.~(\ref{eqna19}) and find for this region a
small deviation for the first overtone as well. Fig.~(\ref{Fig21})
shows a comparison of the numerical result, with the predicted spectrum
for a quadratic dependence of the flux amplitude on the pressure
amplitude.

\subsubsection{Large amplitudes}
For increasing maximum pressure amplitude the deviation factor
$\Gamma$ becomes smaller and finally negative. For really large
pressure amplitude we expect again a significant deviation from the
Goldreich and Wu result.  Fig~(\ref{Fig22}) shows the numerical
result for a $15\%$ pressure amplitude.  As expected from the
analytical result, the difference to the Goldreich and Wu result has
the opposite sign and the spectrum of the $l=2$ mode becomes very
similar to that of an $l=1$ mode. The numerical result for the $l=1$ mode
does not significantly change for an inclination of $60\Deg$ (see
Fig~\ref{Fig17}) and modes with $l=2$ can easily be confused with $l=1$.
\begin{figure}[htb]
\includegraphics*[width=8.8cm]{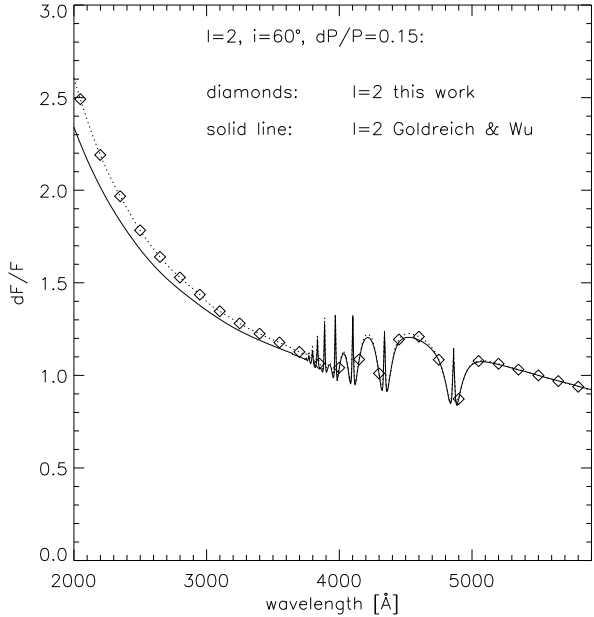}
\caption{Amplitude spectra for the first overtone and large amplitudes.
The lines have the same meaning as in Fig~(\ref{Fig18}).}
\label{Fig23}
\end{figure}

\begin{figure}[htb]
\includegraphics*[width=8.8cm]{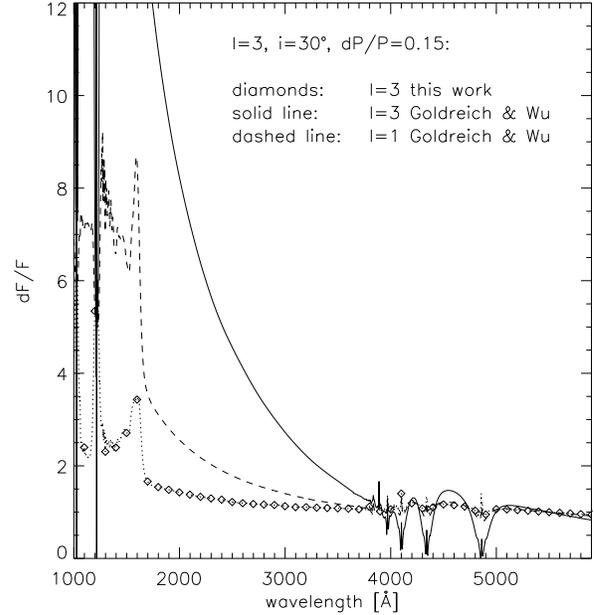}
\caption{Amplitude spectra for $l=3$ and large amplitudes.
The lines have the same meaning as in Fig~\ref{Fig18}.}
\label{Fig24}
\end{figure}

The first overtone is significant for large amplitudes as well. In
contrast to the intermediate amplitudes, a difference from the
predicted result of Goldreich and Wu appears. For our example
Fig.~(\ref{Fig23}) shows the numerical result in comparison with the
quadratic dependence of the flux amplitude.

For $l=3$ the deviation of the numerical results from the results of the
linear assumption is much larger. Fig.~(\ref{Fig24}) shows one
example for a large deviation. In contrast to a rapidly increasing
amplitude in the UV the amplitude remains almost constant. In this
regime the prediction of the variation from Eq.~(\ref{eqna19}) can be
only qualitative. In general the amplitude increase towards the UV is
caused by the decreasing ``effective visible area'' due to the larger
limb darkening. This effect becomes significant, when the effective
area is comparable in size to the spatial scale of the light
variations on the surface of the star. For the large amplitudes the
spatial structure of the $l = 3$ flux variation is more complex than
the unperturbed spherical harmonic and the cancellation effects set in
only farther in the UV.

\section{Conclusions}
In the past the method of time--resolved spectroscopy has been based
on the assumption --- called ``linear'' throughout this paper --- that
the flux varies with the spherical harmonic of the pulsation mode over
the surface of the ZZ~Ceti stars. The numerical light curve simulation
shows that this assumption is questionable in many situations.  For
small amplitudes with sinusoidal lightcurves, only the absolute
amplitude varies like the spherical harmonic of the mode. The phase
shift leads to normalized amplitude spectra, which depend on
inclination and pulsation (pressure) amplitude for mode numbers
$l>2$. For large amplitudes ($>10\%$ in pressure) the absolute
amplitude of the flux depends non--monotonously on the pressure
amplitude and the flux variation shows maxima at different latitudes
than the spherical harmonic of the mode. The numerical results show
relatively good correspondence to traditionally calculated amplitude
spectra for intermediate amplitudes ($\sim 5\%$) in pressure, with
non--sinusoidal light curves. In this regime the phase becomes
stationary and the absolute amplitude varies in good correspondence to
the spherical harmonic of the mode; the numerical results are close to
the results of the perturbation analysis of Goldreich and Wu.

Another result of our simulations is the existence of a maximum
surface flux amplitude. This is a reaction of the convection zone on a
predefined pressure variation in deeper layers, and should not be
confused with the amplitude saturation of the pulsation as studied by
Goldreich \& Wu (\cite{Gold99}). The amplitude of the photospheric
flux is reduced by the inert reaction of the temperature structure in
the convective layer to the changing input flux. The decrease in the
thermal heat flux is converted to kinetic energy of the pulsation as
discussed above, and this convection region is therefore a driving
region for the pulsation. The extent of this conversion depends on the
thermal time scale of the convection zone. 

In our model calculations for large amplitudes the time--averaged
depth of the convection zone becomes much larger than in the
corresponding static model. The thermal time scale then increases with
the pulsation amplitude, leading to a reduction of the surface flux
amplitude. This is the reason for the existence of a maximum amplitude
for the surface flux. It also explains the return to sinusoidal
variations for large pressure amplitudes: the convection zone becomes
so thick that variations during one cycle do not decrease its depth
enough for the appearance of higher overtones.

While our calculations cannot determine the maximum pressure amplitude
that can be reached in a pulsating star, one prediction is that
observable flux amplitudes should not exceed $10\%$, for the
parameters used in this paper. The maximum flux amplitudes support the
dominance of small $l$ in the observed light curves and lead to the
prediction, that the dominant pulsation modes are $l=1$, if $l=1$
modes and modes with larger $l$ are unstable, for a large amplitude
range, independent of the actual amplitudes of the modes.

While nonlinear effects are well known for large observed amplitude
pulsations, the present study reaches the surprising conclusion, that
the amplitudes and phases can be described fairly accurately by the
linear theory as developed by Robinson et al. (\cite{Robi82}) for the
fundamental and Wu (\cite{Wu98}) for higher harmonics.

On the other hand, the observation of small amplitudes in the surface
integrated light does not necessarily guarantee that these effects are
absent. Deviations from the traditional interpretation could occur in
spite of small observed amplitudes under the following conditions
\begin{itemize}
\item[-] the pressure amplitude is larger than about $12\%$, that is
in the range were the flux amplitude decreases again with increasing
pressure amplitude
\item[-] the flux variations are intrinsically large on the surface
but the inclination is close to the latitude of a node line
\item[-] the mode number $l$ is large
\end{itemize}
In all these cases one of the correction factors describing the
difference between our numerical results and the amplitudes calculated
with the linear assumptions becomes large, and mode identification
from time--resolved spectroscopy will only be possible with extended
numerical simulations.

All conclusions are obtained for a special set of stellar and
pulsational parameters. The effects discussed may be more or less
important for other ranges of \Lg\ and \Teff. We have also not taken
into account that in general more than one pulsation mode is present
in a star. The modes will be influenced by the presence of other modes
and by their properties even well below the convection zone. These
effects are beyond the scope of the present analysis.  We do not,
however, expect that the main conclusion will change: the mechanism
that leads to non--sinusoidal light curves for large flux variations
can be important for a correct identification of the pulsation modes
using time--resolved spectroscopy, for any observed amplitude of the
flux.

\begin{acknowledgements}
This work was made possible by a grant from the Deutsche
Forschungsgemeinschaft. We want to thank Dr.~Yanqin Wu, Dr.~Marten van
Kerkwijk, Prof.~Dr.~Rainer Wehrse and Dr.~G\"unther Wuchterl for their
helpful comments and recommendations. D.K. wants to thank
Dr.~J.~Holberg of Lunar and Planetary Lab and Dr.~J.~Liebert of
Steward Observatory for their hospitality during a stay at the
University of Arizona in the summer of 2000 and the DFG for financial
support. 
\end{acknowledgements}

\section*{Appendix A: Difference between the linear spectra and the
numerical results}
To find a semi-analytic expression for the difference between
the synthetic amplitude spectra from the linear theory and
our numerical calculation, we approximate the local Fourier
coefficients $a_1$ and $b_1$ as a quadratic function of the
local pressure amplitude. Assuming a pressure variation
with the spherical harmonic of the mode on the stellar surface
we can write for the modes with $l=0$
\begin{equation}
a_1 \approx \alpha_1\left(\p_{\rm max}\right) \LegP_l
+ \alpha_2\left(\p_{\rm max}\right) (\LegP_l)^2
\end{equation}
The quatities $\alpha_1, \alpha_2$ are given by the fit to the
numerical results and thus depend on the maximum pressure amplitude on
the surface.

In a similar way as for the spherical harmonics we can also expend 
$(\LegP_l)^2$ with a finit number of $\LegP_{l'}$. In this first
order approximation we only need the term with $l'=0$ of
this expansion (the next higher term belongs to $l'=2$ and
is strongly reduced by the cancelation effect). We can write
\begin{equation}
  (\LegP_l)^2 \approx \frac{\LegP_0}{2l+1} = \frac{1}{2l+1}
\end{equation}
Now it is possible to perform the rotation to the coordinate
system of the observer in the space of the spherical harmonic.
The result is a mixing of all possible quantum numbers $m'$
to the same $l$. As in the linear case we only have to
consider the contribution of $m'=0$ to the visible
pulsation amplitude and get
\begin{equation}
  a_1 \approx \alpha_1 \LegP(\cos i)\ \LegP(\cos \vartheta)
  + \alpha_2 \frac{1}{2l+1},
\end{equation}
or after the integration over the visible disc
\begin{equation} 
  A_1 \approx \alpha_1 \LegP(\cos i) \I_l(\lambda,F_0)
  + \alpha_2 \frac{1}{2l+1} \I_0(\lambda,F_0)
\end{equation}
with
\begin{equation}
\I_l \equiv 2\pi \int_0^1 d\cos\vartheta \, \cos\vartheta
  \left. \frac{\partial I}{\partial F} \right|_{F_0}(\cos\vartheta)
  \LegP_l(\cos\vartheta)  \label{defI}
\end{equation}
To find the Fourier amplitude we insert this expression and the
anlogous equation for $B_1$ in (\ref{eqndk1}) and get for small
$\alpha_2,\beta_2$ ($\beta_1$ and $\beta_2$ are the fit parameters for
$B_1$)
\begin{eqnarray}
 \lefteqn{\frac{\delta \flux}{\flux}_1 \approx \frac{2}{A_0}
  \sqrt{(\alpha_1^2+\beta_1^2)(\LegP_l(\cos i)\,\I_l)^2}\cdots }\nonumber \\
  && \mbox{\qquad \qquad} \overline{\cdots + (\alpha_1 \alpha_2+\beta_1
  \beta_2)\frac{\LegP_l(\cos i)\I_0 \I_l}{2l+1}}
\end{eqnarray}
or by expanding the sqare root
\begin{eqnarray} \label{apx1}
 \lefteqn{ \frac{\delta \flux}{\flux}_1  \approx  
\frac{2}{A_0}  \sqrt{\alpha_1^2+\beta_1^2}\ \LegP_l(\cos i)\ \I_l }
\nonumber \\   & & \mbox{\qquad \qquad}
  + \frac{2}{A_0} \frac{\alpha_1\alpha_2 +
  \beta_1\beta_2} {(2l+1)\, \sqrt{\alpha_1^2+\beta_1^2}}\ \I_0
\end{eqnarray}
The first term of this expression is proportional to the result of the linear
theory with the assumption of a flux variation with a spherical harmonic. The
second term is the first order nonlinear correction term for the numerical
result. For this first order approximation it is sufficient to identify the
mean flux $A_0/2$ with the flux of the non-pulsating star $\I_0$. We get for
the correction term of the amplitude
\begin{equation} \label{apx2}
  \frac{\delta \flux}{\flux}_1^{\rm corr} \approx \frac{\alpha_1\alpha_2
  + \beta_1\beta_2}{(2l+1)\sqrt{\alpha_1^2+\beta_1^2}}  
\end{equation}
This expression is a function of $\Teff$ of the equilibrium model, the
pulsation mode and the maximum pulsation amplitude of the model; it
does not depend on the wavelength nor on the inclination. The linear
term, however, which is normally dominant, is a very sensitive
function of the inclination.

Following the standard strategy to find an expression, which
eliminates the inclination dependence of the linear term, we normalize
Eq. (\ref{apx1}) to the amplitude at a reference wavelength
$\lambda_0$ (e.g. $=5500$ \AA) and obtain for the normalized spectrum
\begin{displaymath}
\spec_1 \approx \spec_1^{\rm lin} + \spec_1^{\rm corr} 
\end{displaymath}
with the linear normalized spectrum $\spec_1^{\rm lin}$ and
\begin{equation}
  \spec_1^{\rm corr} \equiv \frac{A_0(\lambda_0)\, \I_0(\lambda)}
  {A_0(\lambda)}
  \ \Gamma\left(\p\right) \Lambda(i,l).
\end{equation}
The inclination dependent term $\Lambda(i,l)$ is given by
\begin{equation}
\Lambda(i,l) \equiv \frac{1}{2l+1} 
\frac{1}{\LegP_l(\cos i)\,\I_l(\lambda_0)}
\end{equation}
and the amplitude dependent term by
\begin{equation}
\Gamma\left(\p\right) \equiv \frac{\alpha_1\alpha_2+\beta_1\beta_2}{
\alpha_1^2 + \beta_1^2}
\end{equation}

\end{document}